\documentclass[pra,twocolumn,superscriptaddress,showpacs]{revtex4}
\usepackage{times}
\usepackage{amsmath}
\usepackage{amsfonts}
\usepackage{amssymb}
\usepackage{graphicx}
\usepackage{pst-all}

\begin{document}

\title{Universal Gates via Fusion and Measurement Operations on SU$(2)_4$ Anyons}

\author{Claire Levaillant}
\affiliation{Department of Mathematics, University of California, Santa Barbara, California 93106, USA}
\author{Bela Bauer}
\affiliation{Station Q, Microsoft Research, Santa Barbara, California 93106-6105, USA}
\author{Michael Freedman}
\affiliation{Station Q, Microsoft Research, Santa Barbara, California 93106-6105, USA}
\affiliation{Department of Mathematics, University of California, Santa Barbara, California 93106, USA}
\author{Zhenghan Wang}
\affiliation{Station Q, Microsoft Research, Santa Barbara, California 93106-6105, USA}
\affiliation{Department of Mathematics, University of California, Santa Barbara, California 93106, USA}
\author{Parsa Bonderson}
\affiliation{Station Q, Microsoft Research, Santa Barbara, California 93106-6105, USA}
\date{\today}

\begin{abstract}
We examine a class of operations for topological quantum computation based on fusing and measuring topological charges for systems with SU$(2)_4$ or $k=4$ Jones-Kauffman anyons. We show that such operations augment the braiding operations, which, by themselves, are not computationally universal. This augmentation results in a computationally universal gate set through the generation of an exact, topologically protected irrational phase gate and an approximate, topologically protected controlled-$Z$ gate.
\end{abstract}

\maketitle









%

\section{Introduction}

The topological approach to quantum computation proposes the strategy of achieving fault-tolerance by utilizing nonlocal state spaces of topological phases of matter, which naturally protect encoded information from local perturbations~\cite{Kitaev03,Freedman98,Preskill98,Freedman02a,Freedman02b,Freedman03b,Nayak08}. In particular, non-Abelian quasiparticles in such a system collectively possess a nonlocal state space which may be used to comprise topologically protected qubits. Topologically protected computational gates acting on such states can be generated through braiding operations obtained by transporting the quasiparticles around each other. Fusion of quasiparticles and topological charge measurement provide a mechanism for initialization and read-out of the computational state, at which point topological protection is explicitly violated in order to extract the information from the topological state space. In Refs.~\cite{Kitaev03,Preskill98,Mochon03,Mochon04}, some additional operations utilizing fusion, measurements, and ancillary anyons were proposed for discrete gauge theories. It was later realized that, by utilizing ancillary anyons, one could generate all braiding transformations (also in a topologically protected manner) using pair-wise topological charge measurements~\cite{Bonderson08a,Bonderson08b} or by adiabatically tuning pair-wise interactions between quasiparticles~\cite{Bonderson12a}.

Recently, the idea of utilizing fusion operations together with topological charge measurements in order to generate topologically protected computational operations has been further explored in Ref.~\cite{Bonderson-unpublished}. In that work, it was demonstrated that such operations could provide useful operations for Ising anyons/Majorana zero modes. These operations augment the computational power of the braiding gates for such quasiparticles, though they are still unable to yield a fully topologically protected computationally universal gate set. More generally, the utility of fusion and measurement operations for topological quantum information processing has been largely unexplored.

In this paper, we advance the study of fusion and measurement operations in topological quantum computation by examining SU$(2)_4$ and $k=4$ Jones-Kauffman (JK$_4$) anyons, which are closely related theories. Potential physical realizations of such anyons may occur as quasiparticles in fractional quantum Hall systems, such as $k=4$ Read-Rezayi states~\cite{Read99} or second level Hermanns hierarchy states~\cite{Hermanns09}, which, respectively, have filling $\nu=2/3$ and $3/5$. Such states provide possible candidates for physically observed Hall plateaus in the second Landau level~\cite{Pan99,Xia04,Pan08,Pan12}.

It is known that braiding transformations of SU$(2)_4$ or JK$_4$ anyons are not computationally universal~\cite{Freedman02b}. We will show that fusion and measurement operations augment the computational power of braiding operations for such anyons, allowing us to produce a computationally universal topologically protected gate set. This is accomplished, in part, through the generation of an exact, topologically protected irrational phase gate. These results provide a compelling demonstration of the utility and value of studying fusion and measurement operations for the purposes of topological quantum computation.

We note that another approach to achieving topologically protected computational universality for SU$(2)_4$ anyons using measurements and non-standard encodings of qutrits was considered in Ref.~\cite{Cui14}.

This paper is structured as follows. In Sec.~\ref{sec:anyon_models}, we provide a brief review of anyon models, topological charge measurements, and some details of the SU$(2)_4$ and JK$_4$ anyon models, upon which the rest of the paper is focused. In Sec.~\ref{sec:Encoding}, we discuss a number of different ways of encoding quantum information in collections of SU$(2)_4$ or JK$_4$ anyonic quasiparticles, and provide protocols for changing between different encodings using fusion and measurement. In Sec.~\ref{sec:braiding_gates}, we present the single qubit computational gates obtained from braiding operations. In Sec.~\ref{sec:X_gate}, we provide a simple fusion-based protocol that generates the NOT gate. In Sec.~\ref{sec:TQF}, we provide a protocol for ``topological qubit fusion,'' which uses fusion and measurements to reduce the number of qubits and collapses the encoded state in a specific nontrivial manner. We explain how this protocol may be used to convert certain ancillary states into computational gates. In Sec.~\ref{sec:irrational}, we provide a protocol for generating a topologically protected irrational phase gate. This is carried out by first utilizing the operations (based on fusion and measurement) developed in previous sections to generate a specific ancillary state, and then utilizing topological qubit fusion to convert the ancillary state into the irrational phase gate. Combined with braiding gates, this provides a computationally universal set of topologically protected single qubit gates. In Sec.~\ref{sec:CZ}, we provide a protocol for generating a topologically protected controlled-$Z$ gate, utilizing the previously developed operations. In Sec.~\ref{sec:discussion}, we make a few concluding remarks. In Appendix~\ref{sec:basic_data}, we give general expressions for the basic data of the SU$(2)_k$ and JK$_k$ anyons models, and tabulate the values of $F$-symbols and $R$-symbols of JK$_4$ anyons that are used in this paper. In Appendix~\ref{sec:random_walk}, we derive the probability of a symmetric 1D random walk completely avoiding negative-valued positions during the first $n$ steps, which is useful for analyzing the implementation of the irrational phase gate.

\section{Anyon Models}
\label{sec:anyon_models}

In this section, we briefly review the basic fusion and braiding properties of quasiparticles (point-like localized excitations) in $(2+1)$D topological phases of matter, as described by anyon models (a.k.a. unitary braided tensor categories). For additional details, see Refs.~\cite{Kitaev06a,Bonderson07b} and references therein.

An anyon model has a finite set $\mathcal{C}$ of topological charges, which obey a commutative, associative fusion algebra
\begin{equation}
a \times b = \sum\limits_{c \in \mathcal{C}} N_{ab}^{c} \, c
,
\end{equation}
where $N_{ab}^{c}$ are positive integers indicating the number of distinct ways charges $a$ and $b$ can be combined to produce charge $c$. There is a unique ``vacuum'' charge, denoted $0$, which has trivial fusion (and braiding) with all other charges (for example, $N_{a 0}^{c} = \delta_{ac}$) and which defines the unique conjugate $\bar{a}$ of each topological charge $a$ via $N_{a b}^{0} = \delta_{\bar{a} b}$.

A quasiparticle in the physical system carries a definite value of topological charge, because it is a localized object. The fusion rules indicate that overall topological charge of a collection of quasiparticles may take superpositions of different values, as long as the quasiparticles are well-separated and there is more than one fusion channel (i.e. $N_{ab}^{c}$ and $N_{ab}^{c'}$ are nonzero for $c \neq c'$). This gives rise to degenerate nonlocal state spaces associated with non-Abelian anyonic quasiparticles, which are topologically protected.

More specifically, each fusion product has an associated vector space $V_{ab}^{c}$ with $\dim{V_{ab}^{c}} = N_{ab}^{c}$, and its dual (splitting) space $V^{ab}_{c}$. The states in these fusion and splitting spaces are assigned to trivalent vertices with the corresponding topological charges. We write orthonormal basis vectors of the fusion and splitting spaces as
\begin{equation}
\left( \frac{d_{c}}{d_{a} d_{b}} \right)^{\frac{1}{4}}
\pspicture[shift=-0.6](-0.1,-0.2)(1.5,-1.2)
  \small
  \psset{linewidth=0.9pt,linecolor=black,arrowscale=1.5,arrowinset=0.15}
  \psline(0.7,0)(0.7,-0.55)
  \psline(0.7,-0.55) (0.25,-1)
  \psline(0.7,-0.55) (1.15,-1)	
  \rput[tl]{0}(0.4,0){$c$}
  \rput[br]{0}(1.4,-0.95){$b$}
  \rput[bl]{0}(0,-0.95){$a$}
 \scriptsize
  \rput[bl]{0}(0.85,-0.5){$\mu$}
  \endpspicture
=\left\langle a,b;c,\mu \right| \in
V_{ab}^{c} ,
\label{eq:bra}
\end{equation}
\begin{equation}
\left( \frac{d_{c}}{d_{a} d_{b}} \right)^{\frac{1}{4}}
\pspicture[shift=-0.6](-0.1,-0.2)(1.5,1.2)
  \small
  \psset{linewidth=0.9pt,linecolor=black,arrowscale=1.5,arrowinset=0.15}
  \psline(0.7,0)(0.7,0.55)
  \psline(0.7,0.55) (0.25,1)
  \psline(0.7,0.55) (1.15,1)	
  \rput[bl]{0}(0.4,0){$c$}
  \rput[br]{0}(1.4,0.8){$b$}
  \rput[bl]{0}(0,0.8){$a$}
 \scriptsize
  \rput[bl]{0}(0.85,0.35){$\mu$}
  \endpspicture
=\left| a,b;c,\mu \right\rangle \in
V_{c}^{ab},
\label{eq:ket}
\end{equation}
where $\mu=1,\ldots ,N_{ab}^{c}$. The normalization factors in terms of $d_{a}$ are included so that the diagrams will be in the isotopy-invariant convention, meaning bending lines and rotating sections of diagrams only change amplitudes by unitary transformations (we refer the reader to Refs.~\cite{Kitaev06a,Bonderson07b} for details). Most anyon models of interest have no fusion multiplicities, i.e. $N_{ab}^{c}=0$ or $1$, in which case the vertex labels $\mu$ are trivial and can be left implicit. We only consider anyon models of this type in this paper, so we will drop these labels from now on. General states and operators are described using fusion/splitting trees constructed by connecting lines with the same topological charge. Charge lines corresponding to quasiparticles have definite values, while charge lines further down the fusion tree, corresponding to nonlocal degrees of freedom, permit superpositions of topological charge values.

In this manner, inner products are formed diagrammatically by stacking the corresponding vertices, which gives
\begin{equation}
\label{eq:inner_product}
  \pspicture[shift=-0.95](-0.2,-0.35)(1.2,1.75)
  \small
  \psarc[linewidth=0.9pt,linecolor=black,border=0pt] (0.8,0.7){0.4}{120}{240}
  \psarc[linewidth=0.9pt,linecolor=black,border=0pt] (0.4,0.7){0.4}{-60}{60}
  \psset{linewidth=0.9pt,linecolor=black,arrowscale=1.5,arrowinset=0.15}
  \psline(0.6,1.05)(0.6,1.55)
  \psline(0.6,-0.15)(0.6,0.35)
  \rput[bl]{0}(0.07,0.55){$a$}
  \rput[bl]{0}(0.94,0.55){$b$}
  \rput[bl]{0}(0.26,1.25){$c$}
  \rput[bl]{0}(0.24,-0.05){$c'$}
  \endpspicture
=\delta _{c c ^{\prime }} \sqrt{\frac{d_{a}d_{b}}{d_{c}}}
  \pspicture[shift=-0.95](0.15,-0.35)(0.8,1.75)
  \small
  \psset{linewidth=0.9pt,linecolor=black,arrowscale=1.5,arrowinset=0.15}
  \psline(0.6,-0.15)(0.6,1.55)
  \rput[bl]{0}(0.75,1.25){$c$}
  \endpspicture
.
\end{equation}%
This relation can be applied inside more complicated diagrams. Note that this diagrammatically encodes charge conservation. Since this formalism describes the states associated with anyonic quasiparticles (in a topological phase of matter), we require the inner product to be positive definite, i.e. $d_a$ are required to be real and positive.

Associativity of fusion is represented in the state space by the $F$-symbols, which (similar to $6j$-symbols) provide a unitary isomorphism relating states written in different bases distinguished by the order of fusion. Diagrammatically, this is represented as
\begin{equation}
\label{eq:F_move}
\psscalebox{.8}{
  \pspicture[shift=-1.0](0,-0.45)(1.8,1.8)
  \small
  \psset{linewidth=0.9pt,linecolor=black,arrowscale=1.5,arrowinset=0.15}
  \psline(0.2,1.5)(1,0.5)
  \psline(1,0.5)(1,0)
  \psline(1.8,1.5) (1,0.5)
  \psline(0.6,1) (1,1.5)
   \rput[bl]{0}(0.05,1.6){$a$}
   \rput[bl]{0}(0.95,1.6){$b$}
   \rput[bl]{0}(1.75,1.6){${c}$}
   \rput[bl]{0}(0.5,0.5){$e$}
   \rput[bl]{0}(0.9,-0.3){$d$}
  \endpspicture
}
= \sum_{f} \left[F_d^{abc}\right]_{ef}
\psscalebox{.8}{
 \pspicture[shift=-1.0](0,-0.45)(1.8,1.8)
  \small
  \psset{linewidth=0.9pt,linecolor=black,arrowscale=1.5,arrowinset=0.15}
  \psline(0.2,1.5)(1,0.5)
  \psline(1,0.5)(1,0)
  \psline(1.8,1.5) (1,0.5)
  \psline(1.4,1) (1,1.5)
   \rput[bl]{0}(0.05,1.6){$a$}
   \rput[bl]{0}(0.95,1.6){$b$}
   \rput[bl]{0}(1.75,1.6){${c}$}
   \rput[bl]{0}(1.25,0.45){$f$}
   \rput[bl]{0}(0.9,-0.3){$d$}
  \endpspicture
}
.
\end{equation}
If the diagram on either side of this equation is prohibited by the fusion rules, the corresponding $F$-symbol is set to $0$.

The quantum dimension of topological charge $a$
\begin{equation}
\label{eq:q_dim}
d_{a} = d_{\bar{a}} \equiv \left| \left[F_a^{a \bar{a} a}\right]_{00} \right|^{-1}
\end{equation}
is also equal to the largest eigenvalue of the matrix ${\bf N}_{a}$ defined by $[{\bf N}_{a}]_{bc} = N^{c}_{ab}$, and so describes how the dimensionality of the state space grows asymptotically as one introduces more quasiparticles of charge $a$ (i.e. dim$[V^{a \ldots a}]\sim d_a ^{n}$ when the number $n$ of charge $a$ quasiparticles is large).
We also define the total quantum dimension of an anyon model to be $\mathcal{D} = \sqrt{ \sum_{a} d_{a}^{2} }$.

The counterclockwise and clockwise braiding exchange operator of topological charges $a$ and $b$ are, respectively, represented diagrammatically as
\begin{equation}
\label{eq:braid}
R^{ab}=
\pspicture[shift=-0.7](-0.1,-0.2)(1.3,1.25)
\small
  \psset{linewidth=0.9pt,linecolor=black,arrowscale=1.5,arrowinset=0.15}
  \psline(0.96,0.05)(0.2,1)
  \psline(0.24,0.05)(1,1)
  \psline[border=2pt](0.24,0.05)(0.92,0.9)
  \rput[bl]{0}(-0.02,1.0){$a$}
  \rput[br]{0}(1.2,1.0){$b$}
  \endpspicture
, \qquad
(R^{ba})^{-1}=
\pspicture[shift=-0.7](-0.1,-0.2)(1.3,1.25)
\small
  \psset{linewidth=0.9pt,linecolor=black,arrowscale=1.5,arrowinset=0.15}
  \psline(0.96,0.05)(0.2,1)
  \psline(0.24,0.05)(1,1)
  \psline[border=2pt](0.96,0.05)(0.28,0.9)
  \rput[bl]{0}(-0.02,1.0){$a$}
  \rput[br]{0}(1.2,1.0){$b$}
  \endpspicture
.
\end{equation}
The action of the braiding operator on the state space can be described in terms of $R$-symbols, which represent the unitary operator for exchanging two anyons in a specific fusion channel, and are obtained by applying the exchange operator to the corresponding trivalent vertices
\begin{equation}
\label{eq:R_move}
\pspicture[shift=-0.65](-0.1,-0.2)(1.5,1.2)
  \small
  \psset{linewidth=0.9pt,linecolor=black,arrowscale=1.5,arrowinset=0.15}
  \psline(0.7,0)(0.7,0.5)
  \psarc(0.8,0.6732051){0.2}{120}{240}
  \psarc(0.6,0.6732051){0.2}{-60}{35}
  \psline (0.6134,0.896410)(0.267,1.09641)
  \psline(0.7,0.846410) (1.1330,1.096410)	
  \rput[bl]{0}(0.4,0){$c$}
  \rput[br]{0}(1.35,0.85){$b$}
  \rput[bl]{0}(0.05,0.85){$a$}
  \endpspicture
= R_{c}^{ab}
\pspicture[shift=-0.65](-0.1,-0.2)(1.5,1.2)
  \small
  \psset{linewidth=0.9pt,linecolor=black,arrowscale=1.5,arrowinset=0.15}
  \psline(0.7,0)(0.7,0.55)
  \psline(0.7,0.55) (0.25,1)
  \psline(0.7,0.55) (1.15,1)	
  \rput[bl]{0}(0.4,0){$c$}
  \rput[br]{0}(1.4,0.8){$b$}
  \rput[bl]{0}(0,0.8){$a$}
  \endpspicture
.
\end{equation}

An anyon model is defined entirely by its $N_{ab}^{c}$, $F$-symbols, and $R$-symbols. The $N_{ab}^{c}$ must provide an associative and commutative algebra. The $F$-symbols and $R$-symbols are constrained by the ``coherence conditions'' (also known as the ``polynomial equations''), which ensure that any two series of $F$ and/or $R$ transformations are equivalent if they start in the same state space and end in the same state space~\cite{MacLane98}. Physically, these consistency conditions are interpreted as enforcing locality in fusion and braiding processes.

An important invariant quantity derived from braiding is the topological twist of charge $a$
\begin{equation}
\theta _{a}=\theta _{\bar{a}}
=\sum\limits_{c,\mu } \frac{d_{c}}{d_{a}}\left[ R_{c}^{aa}\right] _{\mu \mu }
= \frac{1}{d_{a}}
\pspicture[shift=-0.5](-1.3,-0.6)(1.3,0.6)
\small
  \psset{linewidth=0.9pt,linecolor=black,arrowscale=1.5,arrowinset=0.15}
  \psarc[linewidth=0.9pt,linecolor=black] (0.7071,0.0){0.5}{-135}{135}
  \psarc[linewidth=0.9pt,linecolor=black] (-0.7071,0.0){0.5}{45}{315}
  \psline(-0.3536,0.3536)(0.3536,-0.3536)
  \psline[border=2.3pt](-0.3536,-0.3536)(0.3536,0.3536)
  \psline[border=2.3pt](-0.3536,-0.3536)(0.0,0.0)
  \rput[bl]{0}(-0.2,-0.5){$a$}
  \endpspicture
,
\end{equation}
which is a root of unity. Another important invariant quantity is the topological $S$-matrix
\begin{equation}
S_{ab} =\mathcal{D}^{-1}\sum%
\limits_{c}N_{\bar{a} b}^{c}\frac{\theta _{c}}{\theta _{a}\theta _{b}}d_{c}
=\frac{1}{\mathcal{D}}
\pspicture[shift=-0.4](0.0,0.2)(2.6,1.3)
\small
  \psarc[linewidth=0.9pt,linecolor=black] (1.6,0.7){0.5}{167}{373}
  \psarc[linewidth=0.9pt,linecolor=black,border=3pt] (0.9,0.7){0.5}{167}{373}
  \psarc[linewidth=0.9pt,linecolor=black] (0.9,0.7){0.5}{0}{180}
  \psarc[linewidth=0.9pt,linecolor=black,border=3pt] (1.6,0.7){0.5}{45}{150}
  \psarc[linewidth=0.9pt,linecolor=black] (1.6,0.7){0.5}{0}{50}
  \psarc[linewidth=0.9pt,linecolor=black] (1.6,0.7){0.5}{145}{180}
  \rput[bl]{0}(0.2,0.45){$a$}
  \rput[bl]{0}(0.9,0.45){$b$}
  \endpspicture
.
\label{eqn:mtcs}
\end{equation}
When $S$ is unitary, braiding is non-degenerate and the theory is called a ``modular'' theory (MTC).

The projector onto collective topological charge $a$ of $n$ anyons of definite charges $a_1,\ldots,a_n$ is given by
\begin{equation}
\Pi_{a}^{(1 \ldots n)} = \sum_{ c_{2},\ldots,c_{n-1} } \sqrt{\frac{d_{a}}{d_{a_{1}} \ldots d_{a_{n}} }}
\psscalebox{.6}{
 \pspicture[shift=-2](-0.35,-2)(2.5,2.4)
  \small
  \psset{linewidth=0.9pt,linecolor=black,arrowscale=1.5,arrowinset=0.15}
  \psline(0.0,1.75)(1,0.5)
  \psline(2.0,1.75)(1,0.5)
  \psline(0.4,1.25)(0.8,1.75)
   \rput[bl]{0}(-0.15,1.85){$a_1$}
   \rput[bl]{0}(0.75,1.85){$a_2$}
   \rput[bl]{0}(1.95,1.85){$a_n$}
\rput[bl](1.25,1.85){$\cdots$}
\rput{-45}(0.9,1.05){$\cdots$}
   \rput[bl]{0}(0.25,0.65){$c_2$}
  \psset{linewidth=0.9pt,linecolor=black,arrowscale=1.5,arrowinset=0.15}
  \psline(0.0,-1.45)(1,-0.2)
  \psline(2.0,-1.45)(1,-0.2)
  \psline(0.4,-0.95)(0.8,-1.45)
   \rput[bl]{0}(-0.15,-1.75){$a_1$}
   \rput[bl]{0}(0.75,-1.75){$a_2$}
   \rput[bl]{0}(1.95,-1.75){$a_n$}
   \rput[bl]{0}(0.25,-0.6){$c_2$}
   \rput[bl](1.25,-1.75){$\cdots$}
   \rput{45}(0.9,-0.75){$\cdots$}
  \psline(1,-0.2)(1,0.5)
   \rput[bl]{0}(1.15,0.0){$a$}
\endpspicture
}
\label{eq:PIan}
\end{equation}
where we sum over all possible fusion channels (with the same values in the bra as in the ket).

For a modular theory (with unitary $S$-matrix), we can write the projector of $n$ anyons onto definite collective topological charge $a$ by enclosing the charge lines of these anyons with an $\omega_a$ loop
\begin{equation}
\Pi_{a}^{(1 \ldots n)} =
\psscalebox{.8}{
 \pspicture[shift=-2.5](-0.6,-2)(2.8,2.4)
  \small
  \psset{linewidth=0.9pt,linecolor=black,arrowscale=1.5,arrowinset=0.15}
  \psline(0.0,1.75)(0.0,-1.0)
  \psline(2.0,1.75)(2.0,-1.0)
  \psline(0.8,1.75)(0.8,-1.0)
   \rput[bl]{0}(-0.2,1.85){$a_1$}
   \rput[bl]{0}(0.6,1.85){$a_2$}
   \rput[bl]{0}(1.25,1.85){$\ldots$}
   \rput[bl]{0}(1.8,1.85){$a_n$}
  \psellipse[linewidth=0.9pt,linecolor=black,border=0.1](1.0,0.5)(1.5,0.3)
  \psline[linewidth=0.9pt,linecolor=black,border=0.1](0.0,0.5)(0.0,1.0)
  \psline[linewidth=0.9pt,linecolor=black,border=0.1](0.8,0.5)(0.8,1.0)
  \psline[linewidth=0.9pt,linecolor=black,border=0.1](2.0,0.5)(2.0,1.0)
  \psset{linewidth=0.9pt,linecolor=black,arrowscale=1.4,arrowinset=0.15}
  \rput[bl]{0}(2.4,0.15){$\omega_a$}
\endpspicture
}
\label{eq:omega_a_projector}
\end{equation}
where the $\omega_a$-loop is defined as
\begin{equation}
\pspicture[shift=-0.55](-0.25,-0.1)(1.2,1.3)
\small
  \psset{linewidth=0.9pt,linecolor=black,arrowscale=1.5,arrowinset=0.15}
  \psline(0.4,0)(0.4,0.22)
  \psline(0.4,0.45)(0.4,1.2)
  \psellipse[linewidth=0.9pt,linecolor=black,border=0](0.4,0.5)(0.4,0.18)
  \psset{linewidth=0.9pt,linecolor=black,arrowscale=1.4,arrowinset=0.15}
\psline[linewidth=0.9pt,linecolor=black,border=2.5pt](0.4,0.5)(0.4,1.1)
  \rput[bl]{0}(0.8,0.25){$\omega_a$}
  \rput[tl]{0}(0.55,1.2){$b$}
  \endpspicture
= \sum_{x} S_{0a} S_{ax}
\pspicture[shift=-0.55](-0.25,-0.1)(1.2,1.3)
\small
  \psset{linewidth=0.9pt,linecolor=black,arrowscale=1.5,arrowinset=0.15}
  \psline(0.4,0)(0.4,0.22)
  \psline(0.4,0.45)(0.4,1.2)
  \psellipse[linewidth=0.9pt,linecolor=black,border=0](0.4,0.5)(0.4,0.18)
  \psset{linewidth=0.9pt,linecolor=black,arrowscale=1.4,arrowinset=0.15}
\psline[linewidth=0.9pt,linecolor=black,border=2.5pt](0.4,0.5)(0.4,1.1)
  \rput[bl]{0}(0.8,0.25){$x$}
  \rput[tl]{0}(0.55,1.2){$b$}
  \endpspicture
= \delta_{ab}
\pspicture[shift=-0.55](0.05,-0.1)(1,1.3)
\small
  \psset{linewidth=0.9pt,linecolor=black,arrowscale=1.5,arrowinset=0.15}
  \psline(0.4,0)(0.4,1.2)
  \rput[tl]{0}(0.52,1.0){$b$}
  \endpspicture
\label{eq:omega_a_projection}
.
\end{equation}

\subsection{Measurements}

There are two classes of topological charge measurements: local and interferometric.

Local measurements are capable of measuring the topological charge ascribed locally to a single quasiparticle (or similar object) and can also be used to measure the topological charge of the fusion channel of a pair of quasiparticles. Such measurements are physically performed by measuring an externally measurable local quantity, such as energy or fermion parity, which is correlated with the topological charge value of a quasiparticle or small region. In order to measure pairwise fusion channels using such local measurements, one should interpret the diagram of a pairwise projector more physically. Specifically, to perform such a measurement, one must bring the pair of quasiparticles close to each other (or modify the system in some fashion that induces the same effect), essentially fusing them, so that one can apply a local measurement to the fusion outcome of the pair, before finally moving the quasiparticles apart to their original positions (splitting them).

Interferometric measurements of topological charge~\cite{Bonderson07b,Bonderson07c} are physically performed by sending probe quasiparticles through a device which generates coherent superpositions of the probe quasiparticles traveling through distinct paths around some region. The interference between different paths is correlated with the total collective topological charge contained in the encircled region, and this is detectable by measuring the probe quasiparticles exiting the interferometric device. Such measurements are capable of distinguishing between the topological charges ascribed to the overall fusion channel of the entire collection of quasiparticles contained inside the interference loop region of the device. These are non-demolitional measurements of topological charge, in the sense that, as long as the quasiparticles inside the interferometer are kept sufficiently well-separated (distances much larger than the correlation length), only their overall fusion channel is collapsed by such a measurement. State information encoded higher in the fusion tree than the overall fusion channel may be unaffected by the interferometric measurement. We emphasize that this effect cannot be achieved using local topological charge measurements, since bringing more than two quasiparticles close to each other will cause these fusion channels to evolve and decohere in some non-universal manner.

The asymptotic operation of a generically tuned anyonic interferometer using probes that can distinguish all topological charge types will: (1) project the anyonic state onto the subspace where the collective topological charge of everything contained inside the interferometry region (encircled by the probe paths) takes the definite value $a$, and (2) decohere all anyonic entanglement between the interior and exterior regions of the interferometer~\cite{Bonderson07a}. The decohering effect (2) requires the use of density matrix formalism to describe, and it can be described using a superoperator acting on the density matrix (as described in Refs.~\cite{Bonderson07b,Bonderson07c,Bonderson08b}) or, equivalently, by applying $\omega_0$ loops to the density matrix in the diagrammatic representation (as described in Ref.~\cite{Bonderson13b}). It is, however, possible to remove the decohering effect (2) with a protocol that essentially reconnects the severed topological charge line, which is projected onto charge $a$ by the measurement, connecting the interior and exterior regions of the interferometer~\cite{Bonderson-unpublished,Freedman15}. In this manner, we can use interferometry measurements to generate only the topological charge projection (1), which can equivalently be represented by inserting an $\omega_a$ loop around the anyonic charge lines of the quasiparticles in the interior region of the interferometer.

Thus, we can write the effect of both types of topological charge measurements using the standard von Neumann projective measurement formalism applied to state vectors. Specifically, a topological charge measurement for a state $|\Psi \rangle$ with measurement outcome equal to $a$ will occur with probability
\begin{equation}
p_{a} = \langle \Psi | \Pi_{a} |\Psi \rangle = \| \Pi_{a} |\Psi \rangle \|^2
,
\end{equation}
and  transforms to the (normalized) post-measurement state
\begin{equation}
|\Psi \rangle \mapsto \frac{ \Pi_{a} |\Psi \rangle } { \| \Pi_{a} |\Psi \rangle \| }
.
\end{equation}

Throughout this paper, we will always represent measurement projectors using $\omega$-loops, regardless of whether the measurement used is local or interferometric (though when the $\omega$ loop encircles the charge lines of more than two quasiparticles, it necessarily requires an interferometric measurement to achieve).

\subsection{SU$(2)_4$ and JK$_4$ Anyons}

The set of topological charges for SU$(2)_4$ and JK$_4$ anyon models is~\footnote{For SU$(2)_4$ anyons, it is more conventional to use the angular momentum notation, where the topological charge labels are half integers, i.e. $\mathcal{C} = \{ 0,\frac{1}{2},1,\frac{3}{2},2 \}$, but we will be primarily using JK$_4$ anyons in this paper, so we use the integer notation.}
\begin{equation}
\mathcal{C} = \{ 0,1,2,3,4 \}.
\end{equation}
The fusion rules are given by
\begin{equation}
\begin{array}{lll}
0 \times 0 = 0, & 0 \times 1 = 1, & 0 \times 2 = 2, \\
0 \times 3 = 3, & 0 \times 4 = 4, & 1\times 1 = 0 + 2, \\
1\times 2 = 1+3, & 1 \times 3 = 2+4, & 1 \times 4 =3,   \\
2\times 2 = 0 +2+4, & 2 \times 3 = 1+3, & 2 \times 4 =2, \\
3\times 3 = 0 +2 , & 3 \times 4 =1 , & 4 \times 4 =0
\end{array}
\end{equation}
(notice that $\bar{a} = a$ for all $a$) and the quantum dimensions are
\begin{equation}
d_0 = d_4 =1, \quad d_{1}=d_{3} = \sqrt{3}, \quad d_{2} =2
.
\end{equation}
For SU$(2)_{4}$ the topological twist factors are
\begin{equation}
\theta_0 = \theta_4 =1, \quad \theta_{1}=-\theta_{3} = e^{i\frac{\pi}{4}}, \quad \theta_{2} =e^{i\frac{2 \pi}{3}}.
\end{equation}
For JK$_{4}$ the topological twist factors are
\begin{equation}
\theta_0 = \theta_4 =1, \quad \theta_{1}=-\theta_{3} = e^{i\frac{\pi}{4}}, \quad \theta_{2} =e^{-i\frac{2 \pi}{3}}.
\end{equation}

There are two other anyon models possible with these same fusion rules, and they are simply the complex conjugates $\overline{\text{SU}(2)_{4}}$ and $\overline{\text{JK}_{4}}$ of the these two theories (the only difference in the above data being that the twist factors are all complex conjugated).

There are too many $F$-symbols and $R$-symbols to write them all out explicitly, but it is straightforward to compute them using the general expressions given in Appendices~\ref{sec:SU(2)_k} and \ref{sec:JK_k}. We tabulate the $F$-symbols and $R$-symbols for JK$_4$ that we will use for calculations in this paper in Appendix~\ref{sec:JK_4}.

In the following, we focus only on JK$_4$, since these are completely isotopy invariant (all bending and raising and lowering of lines at vertices are done freely)~\footnote{At a more detailed level, being completely isotopy invariant is equivalent to the condition that $\left[ F^{\bar{a}ab}_{b} \right]_{0c} = \left[F^{ab \bar{b}}_{a} \right]_{c0} = \sqrt{\frac{ d_{c} }{d_{a} d_{b} }} $ for all values of $a,b,c$ with $N_{ab}^{c} \neq 0$ (see Ref.~\cite{Bonderson07b}).}, which simplifies the calculations. In particular, this allows us to apply $F$-moves and $R$-moves in any orientation within a diagram. The basic results that we derive for JK$_{4}$ will carry over to SU$(2)_4$ anyons with only minor modifications. The fact that the results carry over follow from the fact that the two theories can be related by simply gluing a semion onto the odd integer charges.~\footnote{More precisely, we have $\text{SU}\left( 2\right) _{4} = \left. \overline{ \text{JK}_{4} } \times \mathbb{Z}_{2}^{(3/2)} \right|_{\mathcal{C}}$, where $\mathbb{Z}_{2}^{(3/2)} = \overline{\text{SU}\left( 2\right) _{1}}$ is a semionic $\mathbb{Z}_{2}$ theory with topological twist $\theta_{1} = -i$, and the charge set of the restricted product is $\mathcal{C} = \{ (0,0),(1,1),(2,0),(3,1),(4,0)\}$.}

\section{Encoding Quantum Information}
\label{sec:Encoding}

There are various ways in which one may encode quantum information in the nonlocal topological state space of non-Abelian anyons. Even when there is a single nontrivial non-Abelian topological charge in the theory, one may choose to encode states more densely or sparsely in the state space.

The ``standard encoding'' of a topological qubit is given by four non-Abelian quasiparticles of the same topological charge $q$ whose collective fusion channel (of the four quasiparticles) is fixed to the vacuum charge $0$, when a pair of charge $q$ quasiparticles has two possible fusion channels. (More generally, if any two of the topological charges $q$ have $d$ possible fusion channels, then this would provide a qudit.)

\subsection{Encoding qubits in JK$_4$ and SU$(2)_4$}

For the JK$_{4}$ and SU$(2)_4$ theories, one may use $q=1$ to form a standard encoding qubit, whose basis states are the two fusion channels $a=0$ and $2$. These are represented diagrammatically as
\begin{equation}
\left| a \right\rangle = \frac{1}{d_1}
\pspicture[shift=*](-1.2,-1)(1.2,0.5)
\psset{unit=0.33cm}\pslabelsep=0.1cm\scriptsize
 \psarc (-2,1){1}{180}{0}
  \psarc (2,1){1}{180}{0}
  \psarc (0,0){2}{180}{0}
  \uput[90] (-3,1){$1$}
  \uput[90] (-1,1){$1$}
  \uput[90] (1,1){$1$}
  \uput[90] (3,1){$1$}
  \uput[-90] (0,-2){$a$}
\endpspicture
\end{equation}
We refer to this as the ``1111'' encoding.

Similarly, the standard encoding using $q=3$ also provides a qubit with fusion channels $a=0$ and $2$. In contrast, the standard encoding using $q=2$ provides a qutrit, whose basis states are the three fusion channels $a=0$, $2$, and $4$.

Another qubit encoding, which will form the primary computational basis utilized in this paper, is given by four quasiparticles, two of which carry topological charge $1$ and two of which carry topological charge $2$, with the collective fusion channel of the four quasiparticles being fixed to vacuum $0$. The topological charges of the two fusion channels encoding the qubit basis states depend on which pair of quasiparticles we fuse. If we fuse a $1$ and a $2$ quasiparticle, the two fusion channels are $a=1$ and $3$. We refer to this last qubit encoding as the ``1221'' encoding, and represent the basis states as
\begin{equation}
\left| a \right\rangle = \frac{1}{\sqrt{ d_1 d_2}}
\pspicture[shift=*](-1.2,-1)(1.2,0.5)
\psset{unit=0.33cm}\pslabelsep=0.1cm\scriptsize
  \psarc (-2,1){1}{180}{0}
  \psarc (2,1){1}{180}{0}
  \psarc (0,0){2}{180}{0}
  \uput[90] (-3,1){$1$}
  \uput[90] (-1,1){$2$}
  \uput[90] (1,1){$2$}
  \uput[90] (3,1){$1$}
  \uput[-90] (0,-2){$a$}
\endpspicture
.
\end{equation}
If we fuse the $1$ and $1$ quasiparticles with each other, or the $2$ and $2$ quasiparticles with each other, the two fusion channels are $a=0$ and $2$. (A pair of charge $2$ quasiparticles may potentially have fusion channels $0$, $2$, and $4$, however, the collective fusion channel restriction of this encoding excludes the charge $4$ fusion channel.) This encoding can be related to the 1221 encoding via the $F$-move
\begin{equation}
\pspicture[shift=*](-1.2,-1)(1.2,0.5)
\psset{unit=0.33cm}\pslabelsep=0.1cm\scriptsize
  \psarc (-2,1){1}{180}{0}
  \psarc (2,1){1}{180}{0}
  \psarc (0,0){2}{180}{0}
  \uput[90] (-3,1){$1$}
  \uput[90] (-1,1){$2$}
  \uput[90] (1,1){$2$}
  \uput[90] (3,1){$1$}
  \uput[-90] (0,-2){$a$}
\endpspicture
= \sum_{b=0,2} \left[ F^{122}_{1} \right]_{ab}
\pspicture[shift=*](-1.2,-1)(1.2,0.5)
\psset{unit=0.33cm}\pslabelsep=0.1cm\scriptsize
  \psarc (0,1){1}{180}{0}
  \psarc (0,1){3}{180}{0}
  \psline(0,0)(0,-2)
  \uput[90] (-3,1){$1$}
  \uput[90] (-1,1){$2$}
  \uput[90] (1,1){$2$}
  \uput[90] (3,1){$1$}
  \uput[-90] (-0.4,-0.5){$b$}
\endpspicture
,
\label{eq:1221_1122}
\end{equation}
so we do not view it as a distinct encoding, but rather as a change of basis.

One may also consider more dense encodings of qubits. In particular, it may be useful to consider two qubits that are encoded in the collective state space of six quasiparticles. For six quasiparticles with respective topological charges 1, 2, 2, 2, 2, and 1, whose combined fusion tree is given by
\begin{equation}
\left| a , b \right\rangle = \frac{1}{d_2 \sqrt{d_1}}
\pspicture[shift=*](-2,-0.75)(2,1)
\psset{unit=0.33cm}\pslabelsep=0.1cm\scriptsize
  \psline (-5,1)(-5,2)
  \psline (-3,1)(-3,2)
  \psline (-1,0)(-1,2)
  \psline (1,0)(1,2)
  \psline (3,1)(3,2)
  \psline (5,1)(5,2)
  \psarc (-4,1){1}{180}{0}
  \psarc (4,1){1}{180}{0}
  \psline (-1,0)(1,0)
  \psarc (-2.5,0){1.5}{180}{0}
  \psarc (2.5,0){1.5}{180}{0}
  \uput[90] (-5,2){$1$}
  \uput[90] (-3,2){$2$}
  \uput[90] (-1,2){$2$}
  \uput[90] (1,2){$2$}
  \uput[90] (3,2){$2$}
  \uput[90] (5,2){$1$}
  \uput[-90] (0,0){$1$}
  \uput[-90] (-2.5,-1.5){$a$}
  \uput[-90] (2.5,-1.5){$b$}
\endpspicture
,
\end{equation}
we have basis states for two qubits given by the possible topological charge values $a=1,3$ and $b=1,3$. We refer to this as the ``two qubit 122221'' encoding.

\subsection{Switching from the 1111 encoding to the 1221 encoding}
\label{sec:encoding_switching}

In order to switch between different encodings of quantum information, one generally needs to utilize operations that involve fusion and measurement. (For a theory in which braiding is dense, one may use braiding operations to change how information is encoded. However, this approach may not be as efficient as the fusion and measurement based approach considered here. Moreover, we are particularly interested in theories for which braiding is not dense.)

We now describe a process that non-deterministically takes one from the 1111 encoding to the 1221 encoding. Starting from a qubit in the 1111 encoding, we follow the steps:
\begin{enumerate}
\item Introduce an ancillary pair of charge $2$ quasiparticles that are pair-produced from vacuum.

\item Fuse one of the charge $1$ quasiparticles with one of the charge $2$ quasiparticles and measure the resulting fusion outcome topological charge $x \in \{ 1,3\}$.

\item Fuse the (new) charge $x$ quasiparticle with one of the original charge $1$ quasiparticles and measure the resulting fusion outcome topological charge $y \in \{ 0,2,4\}$.

\end{enumerate}

The above steps may be applied in many different fashions, e.g. where the ancillary quasiparticles are produced, which quasiparticles are fused, etc. Depending on the specifics of this process, one may need to spatially rearrange the quasiparticles, e.g. via braiding operations, during or after these steps in order to obtain the desired physical processes and final configuration. These details and choices involved in enacting this protocol (following steps 1-3) are most easily specified diagrammatically. For the measurement outcomes $x$ and $y$, we consider the implementation of this protocol specified diagrammatically by:
\begin{equation}
\pspicture[shift=*](-2,-1.2)(3.5,2)
\psset{unit=0.33cm}\pslabelsep=0.1cm\scriptsize
  \pscurve (-2,3)(-0.5,4)(1,3)
  \psframe*[linecolor=white] (-2,2)(1,3)
  \pscurve (0,-3)(2.0,-2.75)(3,-2)
  \psline (-4,1)(-4,5)
  \psline (-2,1)(-2,3)
  \psline (-0.5,4)(-0.5,5)
  \psline (1,2)(1,3)
  \psline (4,1)(4,5)
  \psline (6,1)(6,5)
  \psline[linestyle=dotted] (-7,3)(9,3)
  \psline[linestyle=dotted] (-7,1)(9,1)
  \psarc (-3,1){1}{180}{0}
  \psarc (1,1){1}{0}{180}
  \psarc (3,1){1}{180}{0}
  \psarc (3,1){3}{180}{0}
  \psarc (0,0){3}{180}{-90}
  \uput[90] (-4,5){$1$}
  \uput[90] (-0.5,5){$y$}
  \uput[90] (4,5){$2$}
  \uput[90] (6,5){$1$}
  \uput[0] (1,2.5){$x$}
  \uput[225] (-4,1){$1$}
  \uput[-45] (-2,1){$1$}
  \uput[225] (0,1){$1$}
  \uput[225] (2,1){$2$}
  \uput[-45] (4,1){$2$}
  \uput[225] (6,1){$1$}
  \uput[-90] (0,-3){$a$}
\endpspicture
.
\end{equation}
In this diagram, we have added dotted horizontal lines as a visual aid for partitioning the diagram into three sections corresponding to the three steps of the protocol.

In step 2, the probability of obtaining either fusion outcome $x=1$ or $3$ will be $p_{x}=\frac{1}{2}$. In step 3, the probability of obtaining a particular value of $y$ will, in general, depend on the initial state of the qubit. Thus, the operation in step 3 will read out some state information and collapse the qubit state (at least partially, unless the initial state is in a basis state $|a\rangle$). As usual with measurement processes in quantum mechanics, the state is re-normalized to unit norm.

In order to obtain a state in the 1221 encoding, we require the final fusion outcome to be $y=2$. When $y \neq 2$, the quasiparticles no longer support a degenerate state space, so they can no longer encode a qubit. In this case, we must discard or recycle the quasiparticles. As such, we should only use this protocol on ancillary qubits.

Evaluating the diagram for $y=2$, we obtain the transformation in terms of qubit basis states of the initial 1111 encoding and the final 1221 encoding after fusion and measurement with outcomes $x=1$ or $3$ and $y=2$ to be
\begin{equation}
\pspicture[shift=*](-1.2,-1)(1.2,0.5)
\psset{unit=0.33cm}\pslabelsep=0.1cm\scriptsize
 \psarc (-2,1){1}{180}{0}
  \psarc (2,1){1}{180}{0}
  \psarc (0,0){2}{180}{0}
  \uput[90] (-3,1){$1$}
  \uput[90] (-1,1){$1$}
  \uput[90] (1,1){$1$}
  \uput[90] (3,1){$1$}
  \uput[-90] (0,-2){$a$}
\endpspicture
\mapsto  \sum_{b=1,3} P^{(x)}_{ba}
\pspicture[shift=*](-1.2,-1)(1.2,0.5)
\psset{unit=0.33cm}\pslabelsep=0.1cm\scriptsize
  \psarc (-2,1){1}{180}{0}
  \psarc (2,1){1}{180}{0}
  \psarc (0,0){2}{180}{0}
  \uput[90] (-3,1){$1$}
  \uput[90] (-1,1){$2$}
  \uput[90] (1,1){$2$}
  \uput[90] (3,1){$1$}
  \uput[-90] (0,-2){$b$}
\endpspicture
\end{equation}
where
\begin{equation}
P^{(x)}_{ba} = \frac{1}{\sqrt{d_2}} \left[ F^{x 2 1}_{a} \right]_{1b} \left[ F^{11x}_{b} \right]_{a2}
.
\end{equation}
(Here, we wrote the relation without the normalization constants on the diagrams, and will similarly drop overall constants throughout the paper, whenever they are not important.)
This result is obtained from the following diagrammatic evaluation~\footnote{We note that complete isotopy invariance implies the equivalence of the $F$-symbols $\left[ F^{x 2 1}_{a} \right]_{1b} = \left[ \left(F^{a x 2}_{1} \right)^{-1} \right]_{1b}$, so either can be used in the diagrammatic evaluations.}
\begin{widetext}
\begin{eqnarray}
&& \frac{1}{d_1 \sqrt{d_2}} \left( \frac{d_{x} }{ d_{1} d_{2}  } \right)^{\frac{1}{4}} \left( \frac{d_{2} }{ d_{1} d_{x}  } \right)^{\frac{1}{4}}
\psscalebox{.7}{
\pspicture[shift=*](-2,-1.2)(2.5,2)
\psset{unit=0.33cm}\pslabelsep=0.1cm\scriptsize
  \pscurve (-2,3)(-0.5,4)(1,3)
  \psframe*[linecolor=white] (-2,2)(1,3)
  \pscurve (0,-3)(2.0,-2.75)(3,-2)
  \psline (-4,1)(-4,5)
  \psline (-2,1)(-2,3)
  \psline (-0.5,4)(-0.5,5)
  \psline (1,2)(1,3)
  \psline (4,1)(4,5)
  \psline (6,1)(6,5)
  \psarc (-3,1){1}{180}{0}
  \psarc (1,1){1}{0}{180}
  \psarc (3,1){1}{180}{0}
  \psarc (3,1){3}{180}{0}
  \psarc (0,0){3}{180}{-90}
  \uput[90] (-4,5){$1$}
  \uput[90] (-0.5,5){$2$}
  \uput[90] (4,5){$2$}
  \uput[90] (6,5){$1$}
  \uput[0] (1,2.5){$x$}
  \uput[225] (-4,1){$1$}
  \uput[-45] (-2,1){$1$}
  \uput[225] (0,1){$1$}
  \uput[225] (2,1){$2$}
  \uput[-45] (4,1){$2$}
  \uput[225] (6,1){$1$}
  \uput[-90] (0,-3){$a$}
\endpspicture
}
\quad
=  \frac{1}{ \sqrt{d_1^{3} d_2}}
\pspicture[shift=*](-1.2,-1)(1.2,1)
\psset{unit=0.33cm}\pslabelsep=0.1cm\scriptsize
  \psline (-3,1)(-3,2)
  \psline (-1,1)(-1,2)
  \psline (1,1)(1,2)
  \psline (3,1)(3,2)
  \psarc (-2,1){1}{180}{0}
  \psarc (2,1){1}{180}{0}
  \psline (-1,1)(1,1)
  \psarc (0,0){2}{180}{0}
  \uput[90] (-3,2){$1$}
  \uput[90] (-1,2){$2$}
  \uput[90] (1,2){$2$}
  \uput[90] (3,2){$1$}
  \uput[-90] (0,1){$x$}
  \uput[135] (-1.293,0.293){$1$}
  \uput[45] (1.293,0.293){$1$}
  \uput[-90] (0,-2){$a$}
\endpspicture
\notag
\\
&&= \frac{1}{ \sqrt{d_1^{3} d_2}} \sum_{b} \left[ F^{x 2 1}_{a} \right]_{1b}
\pspicture[shift=*](-1.2,-1)(1.2,1)
\psset{unit=0.33cm}\pslabelsep=0.1cm\scriptsize
  \psline (-3,1)(-3,2)
  \psline (-1,1)(-1,2)
  \psline (1,1)(1,2)
  \psline (3,1)(3,2)
  \psarc (-2,1){1}{180}{0}
  \psarc (2,1){1}{180}{0}
  \pscurve (-1,1)(-0.25,0)(0,-2)
  \psarc (0,0){2}{180}{0}
  \uput[90] (-3,2){$1$}
  \uput[90] (-1,2){$2$}
  \uput[90] (1,2){$2$}
  \uput[90] (3,2){$1$}
  \uput[135] (-1.293,0.293){$1$}
  \uput[45] (-0.25,0){$x$}
  \uput[-120] (-1.732,-1){$a$}
  \uput[-60] (1.732,-1){$b$}
\endpspicture
= \frac{1}{ \sqrt{d_1 d_2}} \sum_{b} \frac{1}{d_1}  \left[ F^{x 2 1}_{a} \right]_{1b} \left[ F^{11x}_{b} \right]_{a2}
\pspicture[shift=*](-1.2,-1)(1.2,1)
\psset{unit=0.33cm}\pslabelsep=0.1cm\scriptsize
  \psline (-3,1)(-3,2)
  \psline (-1,1)(-1,2)
  \psline (1,1)(1,2)
  \psline (3,1)(3,2)
  \psarc (-2,1){1}{180}{0}
  \psarc (2,1){1}{180}{0}
  \psarc (-1,1){0.5}{-104.4775}{90}
  \psarc (0,0){2}{180}{0}
  \uput[90] (-3,2){$1$}
  \uput[90] (-1,2){$2$}
  \uput[90] (1,2){$2$}
  \uput[90] (3,2){$1$}
  \uput[180] (-1,1){$1$}
  \uput[0] (-0.5,1){$x$}
  \uput[-45] (-1.5,0.134){$2$}
  \uput[-90] (0,-2){$b$}
\endpspicture
\notag
\\
&&= \frac{1}{ \sqrt{ d_1 d_2} } \sum_{b} \sqrt{ \frac{d_x }{ d_1 d_2}} \left[ F^{x 2 1}_{a} \right]_{1b} \left[ F^{11x}_{b} \right]_{a2}
\pspicture[shift=*](-1.2,-1)(1.2,0.5)
\psset{unit=0.33cm}\pslabelsep=0.1cm\scriptsize
  \psarc (-2,1){1}{180}{0}
  \psarc (2,1){1}{180}{0}
  \psarc (0,0){2}{180}{0}
  \uput[90] (-3,1){$1$}
  \uput[90] (-1,1){$2$}
  \uput[90] (1,1){$2$}
  \uput[90] (3,1){$1$}
  \uput[-90] (0,-2){$b$}
\endpspicture
\label{eq:encoding_change_step}
\end{eqnarray}
\end{widetext}

We reemphasize that transformations $P^{(x)}$ are not unitary and may (partially) collapse the initial state. Hence, the overall factors here are not important, because one must re-normalize the post-measurement state anyway.
Evaluating the $F$-symbols gives
\begin{equation}
P^{(1)} = \frac{1}{\sqrt{3}} \left[
\begin{array}{cc}
1 & \frac{\sqrt{2}}{4} \\
0 & \frac{3\sqrt{2}}{4}
\end{array}
\right]
, \quad
P^{(3)} = \frac{1}{\sqrt{3}} \left[
\begin{array}{cc}
0 & \frac{3\sqrt{2}}{4} \\
1 & \frac{\sqrt{2}}{4}
\end{array}
\right]
,
\end{equation}
where the columns are assigned values $b=1,3$ and the rows are assigned values $a = 0,2$.

Thus, this implementation of the change of encoding protocol transforms the initial qubit state
\begin{equation}
\left| \Psi \right\rangle = \sum_{a=0,2} \Psi_{a} \left| a \right\rangle
\end{equation}
in the 1111 encoding into the final qubit state
\begin{eqnarray}
\left| \Psi^{\prime}_{x} \right\rangle &=& \frac{ P^{(x)} \left| \Psi \right\rangle }{ \left\|  P^{(x)} \left| \Psi \right\rangle \right\|^{\frac{1}{2} }} \\
P^{(x)} \left| \Psi \right\rangle  &=& \sum\limits_{\substack{ a=0,2 \\ b=1,3}} P^{(x)}_{ba} \Psi_{a} \left| b \right\rangle
\end{eqnarray}
in the 1221 encoding.

Notice that the outcomes $x=1$ and $3$ are related by application of a NOT gate
\begin{equation}
X=
\left[
\begin{array}{cc}
0 & 1 \\
1 & 0
\end{array}
\right]
,
\end{equation}
i.e. $P^{(1)}= X P^{(3)}$. Thus, it is not important which measurement outcome was obtained for $x$ during the protocol, as long as we are able to apply a NOT gate when needed to obtain the desired outcome. In Sec.~\ref{sec:X_gate}, we will describe how to obtain a (deterministic) $X$ gate using a simple fusion process.

We observe that the diagram on the right hand side of the first line of Eq.~(\ref{eq:encoding_change_step}) simply displays the diagrammatic representation of the protocol in a more symmetric form. As such, it emphasizes that different possible protocols (i.e. fusing and measuring in different orders) may be used to produce the same resulting encoding changing operation. We will not go into detail on these, but should keep in mind that alternative realizations may potentially be advantageous.

\subsection{Switching between two qubits in 1221 encodings and a two qubit 122221 encoding}
\label{sec:2x1221_to_122221}

We can switch the encoding of two qubits from two 1221 qubit encodings to a two qubit 122221 encoding, with no state collapse, by using a ``forced measurement'' procedure~\cite{Bonderson08a,Bonderson08b}, as follows. Starting from two 1221 qubits (enumerating the quasiparticles 1-8 from left to right), we follow the steps:

\begin{enumerate}
\item Measure the collective topological charge $x \in \{ 0 , 2 \}$ of quasiparticles 4 and 5. If $x \neq 0$, go to step 2. If $x=0$, go to step 3.

\item Measure the collective topological charge $y \in \{ 0 , 2 \}$ of quasiparticles 5-8. (The outcome is not important.) Go to step 1.

\item Remove the (now ancillary) quasiparticles 4 and 5 in a manner that does not create entanglement, e.g. fuse them together into vacuum or transport them as a pair along the same path away from the other quasiparticles.

\end{enumerate}

Steps 1 and 2 constitute a forced measurement, i.e. a ``repeat until success'' measurement process, which may take $n$ attempts to achieve the desired $x = 0$ measurement outcome. More specifically, the $n$th iteration of the measurement in step 1 is given by
\begin{eqnarray}
&&
\pspicture[shift=*](-2.5,-1)(2.5,1)
\psset{unit=0.33cm}\pslabelsep=0.1cm\scriptsize
  \psline (-7,1)(-7,2)
  \psline (-5,1)(-5,2)
  \psline (-3,1)(-3,2)
  \psline (-1,1)(-1,2)
  \psline (1,1)(1,2)
  \psline (3,1)(3,2)
  \psline (5,1)(5,2)
  \psline (7,1)(7,2)
  \psarc (-6,1){1}{180}{0}
  \psarc (-2,1){1}{180}{0}
  \psarc (2,1){1}{180}{0}
  \psarc (6,1){1}{180}{0}
  \psline (-1,1)(1,1)
  \psarc (-4,0){2}{180}{0}
  \psarc (4,0){2}{180}{0}
  \uput[90] (-7,2){$1$}
  \uput[90] (-5,2){$2$}
  \uput[90] (-3,2){$2$}
  \uput[90] (-1,2){$1$}
  \uput[90] (1,2){$1$}
  \uput[90] (3,2){$2$}
  \uput[90] (5,2){$2$}
  \uput[90] (7,2){$1$}
  \uput[-90] (0,1){$y_n$}
  \uput[-90] (-4,-2){$a$}
  \uput[-90] (4,-2){$b$}
  \uput[180] (-1,0.8){$1$}
  \uput[0] (1,0.8){$1$}
\endpspicture
\notag \\
&& \mapsto
\pspicture[shift=*](-2.5,-1)(2.5,1.5)
\psset{unit=0.33cm}\pslabelsep=0.1cm\scriptsize
  \psline (-7,1)(-7,3)
  \psline (-5,1)(-5,3)
  \psline (-3,1)(-3,3)
  \psline (-1,1)(-1,2)
  \psline (1,1)(1,2)
  \psarc (-1.125,2){0.25}{90}{-90}
  \psarc (1.125,2){0.25}{-90}{90}
  \psline (-1.125,2.25)(1.125,2.25)
  \psline[border=2pt] (-1.125,1.75)(1.125,1.75)
  \psline[border=2pt] (-1,2)(-1,3)
  \psline[border=2pt] (1,2)(1,3)
  \psline (3,1)(3,3)
  \psline (5,1)(5,3)
  \psline (7,1)(7,3)
  \psarc (-6,1){1}{180}{0}
  \psarc (-2,1){1}{180}{0}
  \psarc (2,1){1}{180}{0}
  \psarc (6,1){1}{180}{0}
  \psline (-1,1)(1,1)
  \psarc (-4,0){2}{180}{0}
  \psarc (4,0){2}{180}{0}
  \uput[90] (-7,3){$1$}
  \uput[90] (-5,3){$2$}
  \uput[90] (-3,3){$2$}
  \uput[90] (-1,3){$1$}
  \uput[90] (1,3){$1$}
  \uput[90] (3,3){$2$}
  \uput[90] (5,3){$2$}
  \uput[90] (7,3){$1$}
  \uput[90] (2.1,2.0){$\omega_{x_n}$}
  \uput[-90] (0,1){$y_n$}
  \uput[-90] (-4,-2){$a$}
  \uput[-90] (4,-2){$b$}
  \uput[180] (-1,0.8){$1$}
  \uput[0] (1,0.8){$1$}
\endpspicture
\notag \\
&& = \left[ F^{111}_{1} \right]_{ y_{n} x_{n} }
\pspicture[shift=*](-2.5,-1)(2.5,1.7)
\psset{unit=0.33cm}\pslabelsep=0.1cm\scriptsize
  \psline (-7,1)(-7,4)
  \psline (-5,1)(-5,4)
  \psline (-3,1)(-3,4)
  \psarc (0,4){1}{180}{0}
  \psline (0,2)(0,3)
  \psarc (0,1){1}{0}{180}
  \psline (3,1)(3,4)
  \psline (5,1)(5,4)
  \psline (7,1)(7,4)
  \psarc (-6,1){1}{180}{0}
  \psarc (-2,1){1}{180}{0}
  \psarc (2,1){1}{180}{0}
  \psarc (6,1){1}{180}{0}
  \psarc (-4,0){2}{180}{0}
  \psarc (4,0){2}{180}{0}
  \uput[90] (-7,4){$1$}
  \uput[90] (-5,4){$2$}
  \uput[90] (-3,4){$2$}
  \uput[90] (-1,4){$1$}
  \uput[90] (1,4){$1$}
  \uput[90] (3,4){$2$}
  \uput[90] (5,4){$2$}
  \uput[90] (7,4){$1$}
  \uput[180] (0,2.5){$x_n$}
  \uput[-90] (-4,-2){$a$}
  \uput[-90] (4,-2){$b$}
  \uput[180] (-1,0.8){$1$}
  \uput[0] (1,0.8){$1$}
\endpspicture
\quad
\end{eqnarray}
where $y_1 = 0$, and the $n$th iteration of the measurement in step 2 is
\begin{eqnarray}
&&
\pspicture[shift=*](-2.5,-1)(2.5,1.5)
\psset{unit=0.33cm}\pslabelsep=0.1cm\scriptsize
  \psline (-7,1)(-7,4)
  \psline (-5,1)(-5,4)
  \psline (-3,1)(-3,4)
  \psarc (0,4){1}{180}{0}
  \psline (0,2)(0,3)
  \psarc (0,1){1}{0}{180}
  \psline (3,1)(3,4)
  \psline (5,1)(5,4)
  \psline (7,1)(7,4)
  \psarc (-6,1){1}{180}{0}
  \psarc (-2,1){1}{180}{0}
  \psarc (2,1){1}{180}{0}
  \psarc (6,1){1}{180}{0}
  \psarc (-4,0){2}{180}{0}
  \psarc (4,0){2}{180}{0}
  \uput[90] (-7,4){$1$}
  \uput[90] (-5,4){$2$}
  \uput[90] (-3,4){$2$}
  \uput[90] (-1,4){$1$}
  \uput[90] (1,4){$1$}
  \uput[90] (3,4){$2$}
  \uput[90] (5,4){$2$}
  \uput[90] (7,4){$1$}
  \uput[180] (0,2.5){$x_n$}
  \uput[-90] (-4,-2){$a$}
  \uput[-90] (4,-2){$b$}
  \uput[180] (-1,0.8){$1$}
  \uput[0] (1,0.8){$1$}
\endpspicture
\notag \\
&& \mapsto
\pspicture[shift=*](-2.5,-1)(2.5,2.2)
\psset{unit=0.33cm}\pslabelsep=0.1cm\scriptsize
  \psline (-7,1)(-7,5)
  \psline (-5,1)(-5,5)
  \psline (-3,1)(-3,5)
  \psline (-1,4)(-1,5)
  \psarc (0,4){1}{180}{0}
  \psline (0,2)(0,3)
  \psarc (0,1){1}{0}{180}
  \psline (3,1)(3,4)
  \psline (5,1)(5,4)
  \psline (7,1)(7,4)
  \psarc (0.875,4){0.25}{90}{-90}
  \psarc (7.125,4){0.25}{-90}{90}
  \psline (0.875,4.25)(7.125,4.25)
  \psline[border=2pt] (0.875,3.75)(7.125,3.75)
  \psline[border=2pt] (1,4)(1,5)
  \psline[border=2pt] (3,4)(3,5)
  \psline[border=2pt] (5,4)(5,5)
  \psline[border=2pt] (7,4)(7,5)
  \psarc (-6,1){1}{180}{0}
  \psarc (-2,1){1}{180}{0}
  \psarc (2,1){1}{180}{0}
  \psarc (6,1){1}{180}{0}
  \psarc (-4,0){2}{180}{0}
  \psarc (4,0){2}{180}{0}
   \uput[90] (-7,5){$1$}
  \uput[90] (-5,5){$2$}
  \uput[90] (-3,5){$2$}
  \uput[90] (-1,5){$1$}
  \uput[90] (1,5){$1$}
  \uput[90] (3,5){$2$}
  \uput[90] (5,5){$2$}
  \uput[90] (7,5){$1$}
  \uput[0] (7.375,4){$\omega_{y_{n+1}}$}
  \uput[180] (0,2.5){$x_{n}$}
  \uput[-90] (-4,-2){$a$}
  \uput[-90] (4,-2){$b$}
  \uput[180] (-1,0.8){$1$}
  \uput[0] (1,0.8){$1$}
\endpspicture
\notag \\
&& = \left[ F^{111}_{1} \right]_{ x_{n} y_{n+1} }
\pspicture[shift=*](-2.5,-1)(2.5,1.2)
\psset{unit=0.33cm}\pslabelsep=0.1cm\scriptsize
  \psline (-7,1)(-7,2)
  \psline (-5,1)(-5,2)
  \psline (-3,1)(-3,2)
  \psline (-1,1)(-1,2)
  \psline (1,1)(1,2)
  \psline (3,1)(3,2)
  \psline (5,1)(5,2)
  \psline (7,1)(7,2)
  \psarc (-6,1){1}{180}{0}
  \psarc (-2,1){1}{180}{0}
  \psarc (2,1){1}{180}{0}
  \psarc (6,1){1}{180}{0}
  \psline (-1,1)(1,1)
  \psarc (-4,0){2}{180}{0}
  \psarc (4,0){2}{180}{0}
  \uput[90] (-7,2){$1$}
  \uput[90] (-5,2){$2$}
  \uput[90] (-3,2){$2$}
  \uput[90] (-1,2){$1$}
  \uput[90] (1,2){$1$}
  \uput[90] (3,2){$2$}
  \uput[90] (5,2){$2$}
  \uput[90] (7,2){$1$}
  \uput[-90] (0,1){$y_{n+1}$}
  \uput[-90] (-4,-2){$a$}
  \uput[-90] (4,-2){$b$}
  \uput[180] (-1,0.8){$1$}
  \uput[0] (1,0.8){$1$}
\endpspicture
\quad
\end{eqnarray}

The probability of measurement outcome $x_n = 0$ at the $n$th attempt is
\begin{equation}
p_{0} = \left| \left[ F^{111}_{1} \right]_{ y_{n} 0 } \right|^2 = \frac{d_{y_n}}{ d_{1}^2 } \geq \frac{1}{ d_{1}^2 }= \frac{1}{3}.
\end{equation}
As such, the probability of this process not succeeding within $n$ attempts is at most
\begin{equation}
\text{Prob}(x_1 , \ldots , x_n \neq 0 ) \leq \left(\frac{2}{3} \right)^{n},
\end{equation}
so failure is exponentially suppressed in the number of attempts.
In the notation of Refs.~\cite{Bonderson08a,Bonderson08b}, we write the forced measurement procedure (achieving success at the $n$th attempt) as
\begin{equation}
\breve{\Pi}_{0}^{(45)} = \Pi_{x_n =0}^{(45)} \Pi_{y_n}^{(5678)} \ldots \Pi_{x_1}^{(45)} \Pi_{y_1 =0}^{(5678)}
.
\end{equation}
We emphasize that the measurements made in steps 1 and 2 are independent of the encoded state, as reflected by the independence of the measurement outcome probabilities on $a$ or $b$. This indicates that these measurements do not collapse or otherwise alter the encoded state.

Combining steps 1 and 2 with step 3, which just removes the now ancillary pair of quasiparticles, we obtain the change of encoding
\begin{eqnarray}
&&
\pspicture[shift=*](-2.5,-1)(2.5,1)
\psset{unit=0.33cm}\pslabelsep=0.1cm\scriptsize
  \psline (-7,1)(-7,2)
  \psline (-5,1)(-5,2)
  \psline (-3,1)(-3,2)
  \psline (-1,1)(-1,2)
  \psline (1,1)(1,2)
  \psline (3,1)(3,2)
  \psline (5,1)(5,2)
  \psline (7,1)(7,2)
  \psarc (-6,1){1}{180}{0}
  \psarc (-2,1){1}{180}{0}
  \psarc (2,1){1}{180}{0}
  \psarc (6,1){1}{180}{0}
  \psarc (-4,0){2}{180}{0}
  \psarc (4,0){2}{180}{0}
  \uput[90] (-7,2){$1$}
  \uput[90] (-5,2){$2$}
  \uput[90] (-3,2){$2$}
  \uput[90] (-1,2){$1$}
  \uput[90] (1,2){$1$}
  \uput[90] (3,2){$2$}
  \uput[90] (5,2){$2$}
  \uput[90] (7,2){$1$}
  \uput[-90] (-4,-2){$a$}
  \uput[-90] (4,-2){$b$}
\endpspicture
\notag \\
&& \mapsto
\pspicture[shift=*](-2,-0.75)(2,1)
\psset{unit=0.33cm}\pslabelsep=0.1cm\scriptsize
  \psline (-5,1)(-5,2)
  \psline (-3,1)(-3,2)
  \psline (-1,0)(-1,2)
  \psline (1,0)(1,2)
  \psline (3,1)(3,2)
  \psline (5,1)(5,2)
  \psarc (-4,1){1}{180}{0}
  \psarc (4,1){1}{180}{0}
  \psline (-1,0)(1,0)
  \psarc (-2.5,0){1.5}{180}{0}
  \psarc (2.5,0){1.5}{180}{0}
  \uput[90] (-5,2){$1$}
  \uput[90] (-3,2){$2$}
  \uput[90] (-1,2){$2$}
  \uput[90] (1,2){$2$}
  \uput[90] (3,2){$2$}
  \uput[90] (5,2){$1$}
  \uput[-90] (0,0){$1$}
  \uput[-90] (-2.5,-1.5){$a$}
  \uput[-90] (2.5,-1.5){$b$}
\endpspicture
\end{eqnarray}
which does not collapse or otherwise alter the two qubit state encoded in the fusion channels $a$ and $b$.

It should be clear that the inverse of this change of encoding, taking us from the two qubit 122221 encoding (enumerating the quasiparticles 1-3, 6-8 from left to right, with positions 4 and 5 left vacant for the introduction of ancillary quasiparticles) back to the two 1221 qubits, may be obtained through the steps:
\begin{enumerate}
\item Introduce an ancillary pair of charge 1 quasiparticles that are pair produced from the vacuum, taking the vacant positions of quasiparticles 4 and 5.

\item Measure the collective topological charge $y \in \{ 0 , 2 \}$ of quasiparticles 5-8. (The outcome is not important.) If $y \neq 0$, go to step 3. If $y=0$, stop.

\item Measure the collective topological charge $x \in \{ 0 , 2 \}$ of quasiparticles 4 and 5. Go to step 2.

\end{enumerate}

This utilizes the forced measurement
\begin{equation}
\breve{\Pi}_{0}^{(5678)} =  \Pi_{y_n =0}^{(5678)} \Pi_{x_n }^{(45)} \ldots \Pi_{y_1 }^{(5678)} \Pi_{x_1 =0}^{(45)}
.
\end{equation}

\section{Single qubit braiding gates}
\label{sec:braiding_gates}

Thus far, we have focused only on measurement and fusion operations. We will need to utilize braiding operations as well, so we now present the single qubit braiding gates for the 1111 and 1221 qubit encodings of JK$_{4}$.

\subsection{1111 Single qubit gates}

In the 1111 encoding, we can produce two single qubit gates (and any gates generated by these) by braiding adjacent quasiparticles. These are given by
\begin{equation}
\pspicture[shift=*](-1.2,-1)(1.2,1.5)
\psset{unit=0.33cm}\pslabelsep=0.1cm\scriptsize
  \psarc (-2,4){1}{180}{-150}
  \psarc (-2,4){1}{-30}{0}
  \psline (-1.1375,1.5)(-2.8625,3.5)
  \psline[border=2pt] (-2.8625,1.5)(-1.1375,3.5)
  \psline (1,1)(1,4)
  \psline (3,1)(3,4)
  \psarc (-2,1){1}{150}{30}
  \psarc (2,1){1}{180}{0}
  \psarc (0,0){2}{180}{0}
  \uput[90] (-3,4){$1$}
  \uput[90] (-1,4){$1$}
  \uput[90] (1,4){$1$}
  \uput[90] (3,4){$1$}
  \uput[-90] (0,-2){$a$}
\endpspicture
= R^{11}_{a}
\pspicture[shift=*](-1.2,-1)(1.2,0.5)
\psset{unit=0.33cm}\pslabelsep=0.1cm\scriptsize
 \psarc (-2,1){1}{180}{0}
  \psarc (2,1){1}{180}{0}
  \psarc (0,0){2}{180}{0}
  \uput[90] (-3,1){$1$}
  \uput[90] (-1,1){$1$}
  \uput[90] (1,1){$1$}
  \uput[90] (3,1){$1$}
  \uput[-90] (0,-2){$a$}
\endpspicture
\label{eq:R11}
\end{equation}
and
\begin{eqnarray}
\label{eq:B11}
\pspicture[shift=*](-1.2,-1)(1.2,1.5)
\psset{unit=0.33cm}\pslabelsep=0.1cm\scriptsize
  \psline (-3,1)(-3,4)
  \psarc (0,4){1}{180}{-150}
  \psarc (0,4){1}{-30}{0}
  \psline (0.8625,1.5)(-0.8625,3.5)
  \psline[border=2pt] (-0.8625,1.5)(0.8625,3.5)
  \psarc (0,1){1}{150}{180}
  \psarc (0,1){1}{0}{30}
  \psline (3,1)(3,4)
  \psarc (-2,1){1}{180}{0}
  \psarc (2,1){1}{180}{0}
  \psarc (0,0){2}{180}{0}
  \uput[90] (-3,4){$1$}
  \uput[90] (-1,4){$1$}
  \uput[90] (1,4){$1$}
  \uput[90] (3,4){$1$}
  \uput[-90] (0,-2){$a$}
\endpspicture
&=& \sum_{b=0,2} [B^{111}_{1}]_{ab}
\pspicture[shift=*](-1.2,-1)(1.2,0.5)
\psset{unit=0.33cm}\pslabelsep=0.1cm\scriptsize
 \psarc (-2,1){1}{180}{0}
  \psarc (2,1){1}{180}{0}
  \psarc (0,0){2}{180}{0}
  \uput[90] (-3,1){$1$}
  \uput[90] (-1,1){$1$}
  \uput[90] (1,1){$1$}
  \uput[90] (3,1){$1$}
  \uput[-90] (0,-2){$b$}
\endpspicture
\\
\left[B^{111}_{1}\right]_{ab} &=& \sum_{c=0,2}[F^{111}_{1}]_{ac} R^{11}_{c} [(F^{111}_{1})^{-1}]_{cb}
\quad
\end{eqnarray}

The braiding operation in Eq.~(\ref{eq:R11}) gives the computational gate
\begin{equation}
R_{\pi/3}=
\left[
\begin{array}{cc}
1 & 0 \\
0 & \omega
\end{array}
\right]
\end{equation}
where $\omega = e^{i \frac{2 \pi}{3}}$. The braiding operation in Eq.~(\ref{eq:B11}) gives the gate
\begin{equation}
G=
\frac{i}{\sqrt{3}}\left[
\begin{array}{cc}
1 & \bar{\omega} \sqrt{2} \\
\bar{\omega} \sqrt{2} & -\omega
\end{array}
\right]
.
\end{equation}
The gate set $\{ R_{\pi/3} , G \}$ is not a computationally universal single qubit gate set, since it generates a finite set of 12 gates. This set of gates forms a projective representation of the alternating group $A_4$.

\subsection{1221 Single qubit gates}

In the 1221 encoding, we may also produce unitary single qubit gates by braiding. However, in this case, one cannot use all braiding exchanges of adjacent quasiparticles, because their topological charges are not all identical. We must restrict to the braiding transformations which return the quasiparticles to the initial configuration of charges. The gates that may be obtained from such operations are generated by the following
\begin{equation}
\pspicture[shift=*](-1.2,-1)(1.2,2.5)
\psset{unit=0.33cm}\pslabelsep=0.1cm\scriptsize
  \psarc (-2,7){1}{180}{-150}
  \psarc (-2,7){1}{-30}{0}
  \psline (-1.1375,4.5)(-2.8625,6.5)
  \psline[border=2pt] (-2.8625,4.5)(-1.1375,6.5)
  \psarc (-2,4){1}{150}{-150}
  \psarc (-2,4){1}{-30}{30}
  \psline (-1.1375,1.5)(-2.8625,3.5)
  \psline[border=2pt] (-2.8625,1.5)(-1.1375,3.5)
  \psline (1,1)(1,7)
  \psline (3,1)(3,7)
  \psarc (-2,1){1}{150}{30}
  \psarc (2,1){1}{180}{0}
  \psarc (0,0){2}{180}{0}
  \uput[90] (-3,7){$1$}
  \uput[90] (-1,7){$2$}
  \uput[90] (1,7){$2$}
  \uput[90] (3,7){$1$}
  \uput[-90] (0,-2){$a$}
\endpspicture
= R^{12}_{a} R^{21}_{a}
\pspicture[shift=*](-1.2,-1)(1.2,0.5)
\psset{unit=0.33cm}\pslabelsep=0.1cm\scriptsize
  \psarc (-2,1){1}{180}{0}
  \psarc (2,1){1}{180}{0}
  \psarc (0,0){2}{180}{0}
  \uput[90] (-3,1){$1$}
  \uput[90] (-1,1){$2$}
  \uput[90] (1,1){$2$}
  \uput[90] (3,1){$1$}
  \uput[-90] (0,-2){$a$}
\endpspicture
\label{eq:R12R21}
\end{equation}
and
\begin{eqnarray}
\label{eq:B22}
\pspicture[shift=*](-1.2,-1)(1.2,1.5)
\psset{unit=0.33cm}\pslabelsep=0.1cm\scriptsize
  \psline (-3,1)(-3,4)
  \psarc (0,4){1}{180}{-150}
  \psarc (0,4){1}{-30}{0}
  \psline (0.8625,1.5)(-0.8625,3.5)
  \psline[border=2pt] (-0.8625,1.5)(0.8625,3.5)
  \psarc (0,1){1}{150}{180}
  \psarc (0,1){1}{0}{30}
  \psline (3,1)(3,4)
  \psarc (-2,1){1}{180}{0}
  \psarc (2,1){1}{180}{0}
  \psarc (0,0){2}{180}{0}
  \uput[90] (-3,4){$1$}
  \uput[90] (-1,4){$2$}
  \uput[90] (1,4){$2$}
  \uput[90] (3,4){$1$}
  \uput[-90] (0,-2){$a$}
\endpspicture
&=& \sum_{b=1,3} [B^{122}_{1}]_{ab}
\pspicture[shift=*](-1.2,-1)(1.2,0.5)
\psset{unit=0.33cm}\pslabelsep=0.1cm\scriptsize
  \psarc (-2,1){1}{180}{0}
  \psarc (2,1){1}{180}{0}
  \psarc (0,0){2}{180}{0}
  \uput[90] (-3,1){$1$}
  \uput[90] (-1,1){$2$}
  \uput[90] (1,1){$2$}
  \uput[90] (3,1){$1$}
  \uput[-90] (0,-2){$b$}
\endpspicture
\\
\left[B^{122}_{1}\right]_{ab} &=& \sum_{c = 0,2}[F^{122}_{1}]_{ac} R^{22}_{c} [(F^{122}_{1})^{-1}]_{cb}
\quad
\end{eqnarray}

The pure braid operation in Eq.~(\ref{eq:R12R21}) gives the computational gate
\begin{equation}
Z=
\left[
\begin{array}{cc}
1 & 0 \\
0 & -1
\end{array}
\right]
\end{equation}
and the braiding operation in Eq.~(\ref{eq:B22}) gives the computational gate
\begin{equation}
B=
\frac{1}{2} \left[
\begin{array}{cc}
1 & -i \sqrt{3} \\
-i \sqrt{3} & 1
\end{array}
\right]
.
\end{equation}
We note that exchanging the two charge 1 quasiparticles similarly generates the gate $B$.

The gate set $\{ Z , B \}$ is not a computationally universal single qubit gate set. Indeed, it generates a finite set of 6 gates, which forms a projective representation of the permutation group $S_{3}$ (which is equivalent to the dihedral group $D_{3}$), since $Z^2= -B^3=\openone$ and $BZ = ZB^{-1}$.

\section{Fusion-Based NOT Gate}
\label{sec:X_gate}

We now describe a protocol that generates the NOT gate
\begin{equation}
X=
\left[
\begin{array}{cc}
0 & 1 \\
1 & 0
\end{array}
\right]
\end{equation}
on a 1221 qubit through simple fusion operations that involve no measurements. Starting from a qubit in the 1221 encoding, we follow the steps:

\begin{enumerate}
\item Introduce an ancillary pair of charge 4 quasiparticles that are pair-produced from vacuum.

\item Fuse one of the charge 4 quasiparticles with one of the charge 2 quasiparticles and fuse the other charge 4 quasiparticle with the other charge 2 quasiparticle.

\end{enumerate}

This process is described diagrammatically by
\begin{equation}
\pspicture[shift=*](-2,-1.2)(2,2)
\psset{unit=0.33cm}\pslabelsep=0.1cm\scriptsize
  \psellipse (0,0)(4,3)
  \psframe*[linecolor=white] (-4,0)(4,3.2)
  \psline (-5,1)(-5,5)
  \psline (-3,1)(-3,5)
  \psline (3,1)(3,5)
  \psline (5,1)(5,5)
  \psline[linearc=0.5] (-3,3)(0,1)(3,3)
  \psarc (-4,1){1}{180}{0}
  \psarc (4,1){1}{180}{0}
  \uput[90] (-5,5){$1$}
  \uput[90] (-3,5){$2$}
  \uput[90] (3,5){$2$}
  \uput[90] (5,5){$1$}
  \uput[135] (-5,1){$1$}
  \uput[135] (-3,1){$2$}
  \uput[-90] (-1.5,2){$4$}
  \uput[-90] (1.5,2){$4$}
  \uput[45] (3,1){$2$}
  \uput[45] (5,1){$1$}
  \uput[-90] (0,-3){$a$}
\endpspicture
=
\pspicture[shift=*](-1.2,-1)(1.2,0.5)
\psset{unit=0.33cm}\pslabelsep=0.1cm\scriptsize
  \psarc (-2,1){1}{180}{0}
  \psarc (2,1){1}{180}{0}
  \psarc (0,0){2}{180}{0}
  \uput[90] (-3,1){$1$}
  \uput[90] (-1,1){$2$}
  \uput[90] (1,1){$2$}
  \uput[90] (3,1){$1$}
  \uput[-90] (0,-2){$\neg a$}
\endpspicture
\end{equation}
where $\neg a = 4 \times a = 4-a$ indicates that the effect of this process is the application of a NOT gate. The computational gate resulting from this process is independent of the details of how the charge $4$ quasiparticles are introduced and/or braided around the other quasiparticles, because the charge $4$ quasiparticles are Abelian, and so these details can only result in unimportant overall phases on the state. Moreover, no measurements are necessary (except, perhaps, to ensure that the ancillary quasiparticles indeed carry topological charge 4), because the fusion outcome involving a quasiparticle of charge $4$ is unique, i.e. $4 \times 2 = 2$. We recall that the NOT gate may be used to take us between the encoding changing operations $P^{(1)}$ to $P^{(3)}$ of Sec.~\ref{sec:encoding_switching}.

The gate set $\{ X, Z , B \}$ is not a computationally universal single qubit gate set, either. Indeed, it generates a finite set of 12 gates, which forms a projective representation of the dihedral group $D_{6}$.

\section{Topological Qubit Fusion}
\label{sec:TQF}

Ref.~\cite{Bonderson-unpublished} introduced protocols, referred to as ``topological qubit fusion'' (TQF), which act on multiple topological qubits in a manner that reduces the number of topological qubits through a series of fusion and measurement operations. The reduction of the computational state space using TQF may occur in a nontrivial way, in the sense that it is not simply a projection applied to one of the qubits, but rather may have an effect which is equivalent to applying entangling gates together with measurement projections. TQF was demonstrated to be a useful protocol for converting ancillary states into computational gates~\cite{Bonderson-unpublished}, which is what we will use it for in this paper.

We now describe a TQF process for two qubits in the two qubit 122221 encoding (see also Ref.~\cite{Levaillant15b}), reducing a pair of qubits to a single 1221 qubit. If we wanted to start from two qubits in the 1221 encoding, then the preliminary step would be to apply the protocol of Sec.~\ref{sec:2x1221_to_122221} to switch into the two qubit 122221 encoding.

Starting from two qubits in the two qubit 122221 encoding (enumerating the quasiparticles 1-6 from left to right), we follow the steps:
\begin{enumerate}
\item Measure the collective topological charge $z \in \{ 0,2,4 \}$ of quasiparticles 3 and 4. If $z=2$, go to step 2. If $z=4$, go to step 3. If $z=0$, go to step 4.

\item Measure the collective topological charge $v \in \{ 1,3 \}$ of quasiparticles 4, 5, and 6. Go to step 1.

\item Introduce an ancillary pair of charge 4 quasiparticles. Fuse one of the ancillary charge 4 quasiparticles with quasiparticle 4 and the other one with quasiparticle 5.

\item Remove the (now ancillary) quasiparticles 3 and 4 in a manner that does not create entanglement, e.g. fuse them together into vacuum or transport them as a pair along the same path away from the other quasiparticles.

\end{enumerate}

Steps 1 and 2 constitute a semi-forced (repeat until success) measurement, in the following sense. The measurement outcome $z=2$ is undesired, but, as we will explain, it may be ``undone'' in a manner similar to the forced measurement procedure, since this measurement outcome does not collapse or otherwise alter the encoded state. In contrast, the measurement outcomes $z=0$ and $4$, will collapse the state, as we will also see. The measurement outcome $z=4$ is not necessarily desirable, but it is generally not possible to undo the measurement in this case, since this measurement outcome applies a projection to the encoded state. Thus, we must treat measurement outcomes $z=0$ or $4$ as the ``desired'' outcomes and end the semi-forced measurement process once such an outcome is achieved. This process may take $n$ attempts.

Diagrammatically, the $n$th iteration of step 1 is given by
\begin{eqnarray}
&&
\pspicture[shift=*](-2,-0.75)(2,1)
\psset{unit=0.33cm}\pslabelsep=0.1cm\scriptsize
  \psline (-5,1)(-5,2)
  \psline (-3,1)(-3,2)
  \psline (-1,0)(-1,2)
  \psline (1,0)(1,2)
  \psline (3,1)(3,2)
  \psline (5,1)(5,2)
  \psarc (-4,1){1}{180}{0}
  \psarc (4,1){1}{180}{0}
  \psline (-1,0)(1,0)
  \psarc (-2.5,0){1.5}{180}{0}
  \psarc (2.5,0){1.5}{180}{0}
  \uput[90] (-5,2){$1$}
  \uput[90] (-3,2){$2$}
  \uput[90] (-1,2){$2$}
  \uput[90] (1,2){$2$}
  \uput[90] (3,2){$2$}
  \uput[90] (5,2){$1$}
  \uput[-90] (0,0){$v_{n}$}
  \uput[-90] (-2.5,-1.5){$a$}
  \uput[-90] (2.5,-1.5){$b$}
\endpspicture
\notag \\
&& \mapsto
\pspicture[shift=*](-2,-0.75)(2,1.5)
\psset{unit=0.33cm}\pslabelsep=0.1cm\scriptsize
  \psline (-5,1)(-5,3)
  \psline (-3,1)(-3,3)
  \psline (-1,0)(-1,2)
  \psline (1,0)(1,2)
  \psarc (-1.125,2){0.25}{90}{-90}
  \psarc (1.125,2){0.25}{-90}{90}
  \psline (-1.125,2.25)(1.125,2.25)
  \psline[border=2pt] (-1.125,1.75)(1.125,1.75)
  \psline[border=2pt] (-1,2)(-1,3)
  \psline[border=2pt] (1,2)(1,3)
  \psline (3,1)(3,3)
  \psline (5,1)(5,3)
  \psarc (-4,1){1}{180}{0}
  \psarc (4,1){1}{180}{0}
  \psline (-1,0)(1,0)
  \psarc (-2.5,0){1.5}{180}{0}
  \psarc (2.5,0){1.5}{180}{0}
  \uput[-90] (2.1,1.9){$\omega_{z_n}$}
  \uput[-90] (0,0){$v_n$}
  \uput[90] (-5,3){$1$}
  \uput[90] (-3,3){$2$}
  \uput[90] (-1,3){$2$}
  \uput[90] (1,3){$2$}
  \uput[90] (3,3){$2$}
  \uput[90] (5,3){$1$}
  \uput[-90] (-2.5,-1.5){$a$}
  \uput[-90] (2.5,-1.5){$b$}
\endpspicture
\notag \\
&&= \left[ F^{a22}_{b} \right]_{v_{n} z_{n}}
\pspicture[shift=*](-2,-0.75)(2,1.5)
\psset{unit=0.33cm}\pslabelsep=0.1cm\scriptsize
  \psline (-5,2)(-5,3)
  \psline (-3,2)(-3,3)
  \psarc (0,3){1}{180}{0}
  \psline (0,1)(0,2)
  \psarc (0,0){1}{0}{180}
  \psline (3,2)(3,3)
  \psline (5,2)(5,3)
  \psarc (-4,2){1}{180}{0}
  \psarc (4,2){1}{180}{0}
  \psline (-4,0)(-4,1)
  \psline (4,0)(4,1)
  \psarc (-2.5,0){1.5}{180}{0}
  \psarc (2.5,0){1.5}{180}{0}
  \uput[90] (-5,3){$1$}
  \uput[90] (-3,3){$2$}
  \uput[90] (-1,3){$2$}
  \uput[90] (1,3){$2$}
  \uput[90] (3,3){$2$}
  \uput[90] (5,3){$1$}
  \uput[180] (0,1.5){$z_n$}
  \uput[-90] (-2.5,-1.5){$a$}
  \uput[-90] (2.5,-1.5){$b$}
\endpspicture
\end{eqnarray}
where $v_1 =1$. The probability of measurement outcome $z_{n} =2$ at the $n$th attempt is always $p_{2} = \frac{1}{2}$, independent of the encoded state. This follows from the fact that $\left[ F^{a22}_{b} \right]_{1 2} = \frac{1}{\sqrt{2}}$ and $\left[ F^{a22}_{b} \right]_{3 2} = -\frac{1}{\sqrt{2}}$ for all $a,b \in \{ 1,3 \}$. We emphasize that this verifies our claim that this measurement outcome does not collapse the state encoded in the two qubit degrees of freedom $a$ and $b$. Thus, the probability of achieving the desired result $z_n = 0$ or $4$, is $p_0 + p_4 = \frac{1}{2}$ at each attempt, though we will see that the individual probabilities $p_0$ and $p_4$ will depend on the encoded state. It is easy to see that the fusion rules require that $a=b$ when $z=0$ and that $a=\neg b = 4-b$ when $z=4$ (and that $z=2$ does not impose any additional relation between $a$ and $b$).

The $n$th iteration of step 2 (which is only used when $z_n =2$) is diagrammatically given by
\begin{eqnarray}
&&
\pspicture[shift=*](-2,-0.75)(2,1.5)
\psset{unit=0.33cm}\pslabelsep=0.1cm\scriptsize
  \psline (-5,2)(-5,3)
  \psline (-3,2)(-3,3)
  \psarc (0,3){1}{180}{0}
  \psline (0,1)(0,2)
  \psarc (0,0){1}{0}{180}
  \psline (3,2)(3,3)
  \psline (5,2)(5,3)
  \psarc (-4,2){1}{180}{0}
  \psarc (4,2){1}{180}{0}
  \psline (-4,0)(-4,1)
  \psline (4,0)(4,1)
  \psarc (-2.5,0){1.5}{180}{0}
  \psarc (2.5,0){1.5}{180}{0}
  \uput[90] (-5,3){$1$}
  \uput[90] (-3,3){$2$}
  \uput[90] (-1,3){$2$}
  \uput[90] (1,3){$2$}
  \uput[90] (3,3){$2$}
  \uput[90] (5,3){$1$}
  \uput[180] (0,1.5){$2$}
  \uput[-90] (-2.5,-1.5){$a$}
  \uput[-90] (2.5,-1.5){$b$}
\endpspicture
\notag \\
&& \mapsto
\pspicture[shift=*](-2,-0.75)(2,1.7)
\psset{unit=0.33cm}\pslabelsep=0.1cm\scriptsize
  \psline (-5,2)(-5,4)
  \psline (-3,2)(-3,4)
  \psarc (0,3){1}{180}{0}
  \psline (0,1)(0,2)
  \psarc (0,0){1}{0}{180}
  \psline (3,2)(3,3)
  \psline (5,2)(5,3)
  \psarc (0.875,3){0.25}{90}{-90}
  \psarc (5.125,3){0.25}{-90}{90}
  \psline (0.875,3.25)(5.125,3.25)
  \psline (-1,3)(-1,4)
  \psline[border=2pt] (0.875,2.75)(5.125,2.75)
  \psline[border=2pt] (1,3)(1,4)
  \psline[border=2pt] (3,3)(3,4)
  \psline[border=2pt] (5,3)(5,4)
  \psarc (-4,2){1}{180}{0}
  \psarc (4,2){1}{180}{0}
  \psline (-4,0)(-4,1)
  \psline (4,0)(4,1)
  \psarc (-2.5,0){1.5}{180}{0}
  \psarc (2.5,0){1.5}{180}{0}
  \uput[90] (-5,4){$1$}
  \uput[90] (-3,4){$2$}
  \uput[90] (-1,4){$2$}
  \uput[90] (1,4){$2$}
  \uput[90] (3,4){$2$}
  \uput[90] (5,4){$1$}
  \uput[0] (5.375,3){$\omega_{v_{n+1}}$}
  \uput[180] (0,1.5){$2$}
  \uput[-90] (-2.5,-1.5){$a$}
  \uput[-90] (2.5,-1.5){$b$}
\endpspicture
\notag \\
&&= \left[ F^{22b}_{a} \right]_{2 v_{n+1}}
\pspicture[shift=*](-2,-0.75)(2,1.1)
\psset{unit=0.33cm}\pslabelsep=0.1cm\scriptsize
  \psline (-5,1)(-5,2)
  \psline (-3,1)(-3,2)
  \psline (-1,0)(-1,2)
  \psline (1,0)(1,2)
  \psline (3,1)(3,2)
  \psline (5,1)(5,2)
  \psarc (-4,1){1}{180}{0}
  \psarc (4,1){1}{180}{0}
  \psline (-1,0)(1,0)
  \psarc (-2.5,0){1.5}{180}{0}
  \psarc (2.5,0){1.5}{180}{0}
  \uput[90] (-5,2){$1$}
  \uput[90] (-3,2){$2$}
  \uput[90] (-1,2){$2$}
  \uput[90] (1,2){$2$}
  \uput[90] (3,2){$2$}
  \uput[90] (5,2){$1$}
  \uput[-90] (0,0){$v_{n+1}$}
  \uput[-90] (-2.5,-1.5){$a$}
  \uput[-90] (2.5,-1.5){$b$}
\endpspicture
.
\end{eqnarray}
Again, the outcome is independent of the encoded state and does not collapse nor alter it.

Step 3 simply decouples the quasiparticles 3 and 4 from the other computational quasiparticles when the measurement outcome is $z=4$.
The process in step 3 (which is only used when $z_n =4$) is represented diagrammatically by
\begin{eqnarray}
&&
\pspicture[shift=*](-2,-0.75)(2,1.8)
\psset{unit=0.33cm}\pslabelsep=0.1cm\scriptsize
  \psline (-5,2)(-5,4)
  \psline (-3,2)(-3,4)
  \psarc (0,3){1}{180}{0}
  \psline (0,1)(0,2)
  \psarc (0,0){1}{0}{180}
  \psline (3,2)(3,4)
  \psline (5,2)(5,4)
  \psarc (-4,2){1}{180}{0}
  \psarc (4,2){1}{180}{0}
  \psline (-4,0)(-4,1)
  \psline (4,0)(4,1)
  \psarc (-2.5,0){1.5}{180}{0}
  \psarc (2.5,0){1.5}{180}{0}
  \psline (-1,3)(-1,4)
  \psline (1,3)(1,4)
  \psline[linearc=0.5] (1,3)(2,2)(3,3)
  \uput[90] (-5,4){$1$}
  \uput[90] (-3,4){$2$}
  \uput[90] (-1,4){$2$}
  \uput[90] (1,4){$2$}
  \uput[90] (3,4){$2$}
  \uput[90] (5,4){$1$}
  \uput[180] (0,1.5){$4$}
  \uput[-90] (-2.5,-1.5){$a$}
  \uput[-90] (2.5,-1.5){$b$}
  \uput[-90] (2,2.3){$4$}
\endpspicture
=
\pspicture[shift=*](-2,-0.75)(2,1.5)
\psset{unit=0.33cm}\pslabelsep=0.1cm\scriptsize
  \psline (-5,2)(-5,3)
  \psline (-3,2)(-3,3)
  \psarc (0,3){1}{180}{0}
  \psarc (0,0){1}{0}{180}
  \psline (3,2)(3,3)
  \psline (5,2)(5,3)
  \psarc (-4,2){1}{180}{0}
  \psarc (4,2){1}{180}{0}
  \psline (-4,0)(-4,1)
  \psline (4,0)(4,1)
  \psarc (-2.5,0){1.5}{180}{0}
  \psarc (2.5,0){1.5}{180}{0}
  \psline (1,0)(3,2)
  \uput[90] (1.8,1){$4$}
  \uput[90] (-5,3){$1$}
  \uput[90] (-3,3){$2$}
  \uput[90] (-1,3){$2$}
  \uput[90] (1,3){$2$}
  \uput[90] (3,3){$2$}
  \uput[90] (5,3){$1$}
  \uput[-90] (-2.5,-1.5){$a$}
  \uput[-90] (2.5,-1.5){$b$}
\endpspicture
\notag \\
&& \qquad \qquad = \left[ F^{421}_{b} \right]_{2 a}
\pspicture[shift=*](-2,-0.75)(2,1.5)
\psset{unit=0.33cm}\pslabelsep=0.1cm\scriptsize
  \psline (-5,2)(-5,3)
  \psline (-3,2)(-3,3)
  \psarc (0,3){1}{180}{0}
  \psarc (0,0){1}{0}{180}
  \psline (3,2)(3,3)
  \psline (5,2)(5,3)
  \psarc (-4,2){1}{180}{0}
  \psarc (4,2){1}{180}{0}
  \psline (-4,0)(-4,1)
  \psline (4,0)(4,1)
  \psarc (-2.5,0){1.5}{180}{0}
  \psarc (2.5,0){1.5}{180}{0}
  \uput[90] (-5,3){$1$}
  \uput[90] (-3,3){$2$}
  \uput[90] (-1,3){$2$}
  \uput[90] (1,3){$2$}
  \uput[90] (3,3){$2$}
  \uput[90] (5,3){$1$}
  \uput[-90] (-2.5,-1.5){$a$}
\endpspicture
.
\end{eqnarray}

Step 4 removes the now ancillary quasiparticles 3 and 4, leaving a single qubit in the 1221 encoding.

We denote the operation corresponding to the TQF process described in this section, with final measurement outcome $z=0$ or $4$, by $Q^{(z)}$, with the effect on the two qubit basis states given by
\begin{equation}
|a,b \rangle \mapsto Q^{(z)} |a,b \rangle = \sum_{c=1,3} Q^{(z)}_{c;a,b}|c \rangle
\end{equation}
where
\begin{equation}
Q^{(z)}_{c;a,b} = \left[ F^{a22}_{b} \right]_{v z} \left[ F^{z21}_{b} \right]_{2 a} \delta_{a c}
.
\end{equation}
Evaluating the $F$-symbols, we have
\begin{equation}
Q^{(z)}_{c; a,b} = \frac{1}{\sqrt{2}} \delta_{a, b \times z} \delta_{a c}
,
\end{equation}
which we can write in matrix notation as
\begin{equation}
Q^{(0)} = \left[
\begin{array}{cccc}
\frac{1}{\sqrt{2}} & 0 & 0 & 0 \\
0 & 0 & 0 & \frac{1}{\sqrt{2}}
\end{array}
\right],
\quad
Q^{(4)} = \left[
\begin{array}{cccc}
0 & \frac{1}{\sqrt{2}} & 0 & 0 \\
0 & 0 & \frac{1}{\sqrt{2}} & 0
\end{array}
\right]
\end{equation}

Thus, for a two qubit state $|\Psi \rangle = \sum_{a,b=1,3} \Psi_{a,b} |a,b \rangle$, the effect of the TQF process with measurement outcomes $z=0$ and $4$ are, respectively, given by the maps
\begin{eqnarray}
|\Psi \rangle   &\mapsto&  \frac{Q^{(0)}|\Psi \rangle}{\| Q^{(0)}|\Psi \rangle \|} = \frac{\Psi_{1,1} |1 \rangle + \Psi_{3,3} |3 \rangle }{\sqrt{ \left|\Psi_{1,1} \right|^2 +\left|\Psi_{3,3} \right|^2 } }   \\
|\Psi \rangle   &\mapsto&  \frac{Q^{(4)}|\Psi \rangle}{\| Q^{(4)}|\Psi \rangle \|} = \frac{\Psi_{1,3} |1 \rangle + \Psi_{3,1} |3 \rangle }{\sqrt{ \left|\Psi_{1,3} \right|^2 +\left|\Psi_{3,1} \right|^2 } }
\end{eqnarray}
to single qubit states. The possible outcomes $z=0$ and $4$ of the entire TQF process will, respectively, occur with probabilities $p_0 = |\Psi_{1,1}|^2 + |\Psi_{3,3}|^2$ and $p_4 = |\Psi_{1,3}|^2 + |\Psi_{3,1}|^2$.
We emphasize that this TQF process is not deterministic, as the probability of these outcomes generally depend on the quantum state encoded in the qubits. As such, they generically collapse the initial encoded state (in some nontrivial fashion), so one should be careful how and when the TQF process is used.

\subsection{Converting states into gates}
\label{sec:Converting}

As demonstrated in Ref.~\cite{Bonderson-unpublished}, one of the useful applications of TQF is to convert ancillary states into operations on a computational state. If we use an ancillary qubit in a state with equal magnitude superposition of its basis states, this will generate a unitary phase gate on the computational qubit. If we use a non-balanced ancillary qubit state, the effect on the computational state will generically not be unitary, since it will involve some projection.

We now describe this protocol for JK$_{4}$ anyons with qubits in the 1221 encoding and the ancillary qubit in the state
\begin{equation}
| R_{\phi /2} \rangle = \frac{1}{\sqrt{2}} \left( e^{-i \frac{\phi}{2} } | 1 \rangle + e^{i \frac{\phi}{2} } | 3 \rangle \right)
.
\end{equation}

Starting from a computational state $| \Psi \rangle$ with a qubit (at least one) in the 1221 encoding and an ancillary qubit state $| R_{\phi /2} \rangle$ in the 1221 encoding, we follow the steps:
\begin{enumerate}
\item Switch the two 1221 qubits (computational and ancillary) into the two qubit 122221 encoding.

\item Apply TQF to these two qubits.

\end{enumerate}

This protocol will have TQF outcome $z=0$ and $4$ with equal probability $p_0 = p_4 =\frac{1}{2}$. Using the analysis of our TQF operation, the resulting operations for these two outcomes are, respectively
\begin{eqnarray}
|\Psi \rangle | R_{\phi /2} \rangle   &\mapsto&  \sqrt{2} Q^{(0)}|\Psi \rangle | R_{\phi /2} \rangle  = R_{\phi /2} |\Psi \rangle   \\
|\Psi \rangle | R_{\phi /2} \rangle   &\mapsto&  \sqrt{2} Q^{(4)}|\Psi \rangle | R_{\phi /2} \rangle  = R_{- \phi /2} |\Psi \rangle
.
\end{eqnarray}
We emphasize that this protocol essentially consumes the ancillary qubit in order to convert it into the phase gate
\begin{equation}
R_{\pm \phi /2} = \left[
\begin{array}{cc}
1 & 0 \\
0 & e^{\pm i \phi }
\end{array}
\right]
\end{equation}
acting on the computational state, where the $+$ and $-$ phases are obtained with equal probabilities.

\section{Irrational Phase Gate}
\label{sec:irrational}

The topologically protected computational gates obtained from braiding transformations in the 1221 qubit encoding, which are generated by the gate set $\{ Z , B \}$, are not computationally universal. (Supplementing the braiding gates with the gate $X$ does not make them computationally universal, either.) We propose to supplement these operations with an irrational phase gate
\begin{equation}
K= R_{\alpha/2} =
\left[
\begin{array}{cc}
1 & 0 \\
0 & e^{i \alpha}
\end{array}
\right]
,
\end{equation}
where $e^{i \alpha} = \frac{-1 + i 4\sqrt{3}}{7}$. Olmsted's theorem~\cite{Olmsted1945} assures us that $\alpha/2\pi$ is an irrational number~\footnote{Olmsted's theorem states that, if $x$ is rational in degrees, then the only possible rational values of $\sin x$ or $\cos x$ are $0, \pm \frac{1}{2},\pm 1$.} and, hence, $K$ has infinite order.

It is straightforward to demonstrate that the gate set $\{ X, Z, B, K \}$ is computationally universal for single qubits. In order to verify that any given set of single qubit gates is computationally universal (dense in the set of all single qubit gates), we can use the fact that the only infinite proper Lie subgroups of $\text{PU}(2) \cong \text{SO}(3)$ are isomorphic to either $\text{U}(1) \cong \text{SO}(2)$ or to O$(2)$. It follows that, if an infinite subgroup of PU$(2)$ is not isomorphic to a subgroup of either U$(1)$ or O$(2)$, then its closure must be equal to PU$(2)$. Clearly, the gates $B$ and $K$ projectively generate an infinite subgroup of PU$(2)$. However, $B$ and $K$ do not (projectively) commute with each other, so they cannot generate a subgroup isomorphic to U$(1)$. Additionally, we notice that the (nested) group commutator $[[B,K^{-1}], [B,K]] \neq \openone$, where the group commutator $[a,b]$ is defined here to be $a^{-1}b^{-1}ab$. (There is no distinction between checking this simply by multiplying matrices and checking it in the projective quotient SO$(3)$, since overall phases cancel inside a commutator.) This implies the infinite subgroup generated by $B$ and $K$ cannot be a subgroup of any $2$-stage solvable group, such as O$(2)$. Thus, the gate set $\{B,K \}$ must generate a subgroup of PU$(2)$ that is dense, i.e. has closure equal to PU$(2)$.

We have shown in Sec.~\ref{sec:Converting} how to convert an ancillary qubit in the state $| R_{\phi /2} \rangle$ into a unitary phase gate $R_{\pm \phi /2}$, so what remains to be shown is that we can generate the irrational state
\begin{eqnarray}
| K \rangle &=& \sqrt{\frac{3}{14}} \left[ \left(1 - i \frac{2 \sqrt{3} }{3} \right) | 1 \rangle + \left(1 + i \frac{2 \sqrt{3} }{3} \right) | 3 \rangle \right]  \notag \\
&=& \frac{1}{\sqrt{2}} \left(  e^{-i \frac{\alpha}{2} }  | 1 \rangle + e^{i \frac{\alpha}{2} } | 3 \rangle \right)
= \left| R_{\alpha /2} \right\rangle
\end{eqnarray}
in the 1221 encoding, using measurements, braiding, and fusion operations. The previous sections have developed all the operational tools we need to produce this state, so it only remains to assemble them.

\subsection{Generating the irrational state $|K\rangle$}

As a preliminary step in producing an ancillary 1221 qubit in the irrational state $|K\rangle$, we first prepare two ancillary 1221 qubits in the states
\begin{eqnarray}
\label{eq:Phi_{+1,1}}
| \Phi_{+1,1} \rangle &=& \sqrt{\frac{3}{10}} \left[ \left(1 - i \frac{2 \sqrt{3} }{3} \right) | 1 \rangle + | 3 \rangle \right] \\
| \Phi_{-1,3} \rangle &=& \sqrt{\frac{3}{10}} \left[  | 1 \rangle + \left( 1 +i \frac{2 \sqrt{3} }{3} \right) | 3 \rangle \right]
.
\label{eq:Phi_{-1,3}}
\end{eqnarray}
We generate ancillary qubits in these states using the following protocol.

Starting from an ancillary qubit in the 1111 encoding (enumerating the quasiparticle 1-4), we follow the steps:
\begin{enumerate}
\item Initialize the 1111 qubit in the state $| 2 \rangle$.

\item Perform a braiding operation interchanging quasiparticles 2 and 3.

\item Switch from the 1111 encoding to the 1221 encoding, using the protocol of Sec.~\ref{sec:encoding_switching}.

\end{enumerate}

Step 1 may be carried out in various different ways. For example, one may initialize the qubit in the state $| 2 \rangle$ using a forced measurement~\cite{Bonderson08a,Bonderson08b}. Specifically, this means first measure the collective topological charge of quasiparticles 2 and 3 (the outcome $0$ or $2$ is unimportant) and then measure the collective topological charge of quasiparticles 1 and 2. If the outcome of the second measurement is charge $2$, then the desired state has been prepared. If the outcome of the second measurement is charge $0$, then simply repeat these two measurements until the outcome of the second measurement is $2$. Another way to obtain this initialized state is given by replacing the measurements of quasiparticles 2 and 3 in this forced measurement procedure with braiding operations of quasiparticles 2 and 3.

In step 2, the braiding operation may be counterclockwise or clockwise, which we will distinguish with a label $s=+1$ or $-1$, respectively. This applies the gate $G^{s}$ to the qubit.

In step 3, the encoding change involves a measurement outcome $x=1$ or $3$, as described in Sec.~\ref{sec:encoding_switching}. If the desired measurement outcome $x$ is not obtained, we can always apply a NOT gate to switch to the desired outcome, as previously explained. There is also a second measurement in this step, which is assumed to be $y=2$, in order to result in a state in the 1221 encoding. When this measurement yields $y \neq 2$, we must discard or recycle the quasiparticles and start over. This is not a problem, since we are only generating ancillary states at this stage.

The resulting ancillary qubit produced from this protocol is in the 1221 encoding and in the state
\begin{equation}
| \Phi_{s,x} \rangle = \frac{ P^{(x)} G^{s} | 2 \rangle }{ \| P^{(x)} G^{s} | 2 \rangle \|}
.
\end{equation}
It is straightforward to check that this gives the states in Eqs.~(\ref{eq:Phi_{+1,1}}) and (\ref{eq:Phi_{-1,3}}).

Finally, in order to obtain an ancillary 1221 qubit in the irrational state $|K\rangle$, we start from two ancillary 1221 qubits, respectively prepared in the states $| \Phi_{+1,1} \rangle$ and $| \Phi_{-1,3} \rangle$, and follow the steps:
\begin{enumerate}
\item Switch the two 1221 qubits into the two qubit 122221 encoding.

\item Apply TQF to these two qubits.

\end{enumerate}
{F}rom the analysis of Sec.~\ref{sec:TQF}, we find that the result of the TQF operation on this pair of ancillary qubits with measurement outcome $z=0$ is the desired ancillary 1221 qubit state
\begin{equation}
| \Phi_{+1,1} \rangle | \Phi_{-1,3} \rangle \mapsto \frac{Q^{(0)}  | \Phi_{+1,1} \rangle  | \Phi_{-1,3} \rangle }{ \| Q^{(0)} | \Phi_{+1,1} \rangle  | \Phi_{-1,3} \rangle  \|} = | K \rangle
.
\end{equation}
When the TQF operation has measurement outcome $z=4$, the resulting state is not in the desired form, nor can it easily be salvaged for our purposes. Since these are ancillary qubits, we may simply discard or recycle the final qubit when the TQF operation has $z=4$.

\subsection{Converting $|K\rangle$ into the phase gate $K$}

Once we have an ancillary qubit in the 1221 encoding prepared in the irrational state $| K \rangle$, we can use TQF to convert the ancillary qubit into a phase gate acting on a computational qubit in the 1221 encoding, following the protocol described in Sec.~\ref{sec:Converting}. However, this is not a deterministic process, as the measurement outcomes $z=0$ and $4$ of the TQF procedure will occur with equal probabilities $p_0 = p_4 =\frac{1}{2}$. The outcome $z=0$ results in an application of the unitary phase gate $K$ to the computational state, while the outcome $z=4$ results in an application of $K^{-1}$. If we intended to apply a $K$ gate in our quantum computation, but instead generated the gate $K^{-1}$ from this protocol, we need a strategy for correcting this undesired outcome. We obviously cannot correct this simply by applying gates from our deterministic gate set $\{ X , Z , B \}$.

A simple strategy for dealing with this issue is to repeatedly apply the protocol for converting states $|K\rangle$ into phase gates until the product of gates applied is equal to the desired phase gate $K$. More specifically, after one application of the protocol, we will generate the $K$ gate with probability $1/2$ and $K^{-1}$ with probability $1/2$. If we generated the undesired gate $K^{-1}$, then we apply the protocol two more times. These two applications have probability $1/4$ of generating $K^2$, probability $1/2$ of generating $\openone$, and probability $1/4$ of generating $K^{-2}$. Thus, following the initial gate $K^{-1}$, the two additional applications of the protocol have probability $1/4$ of making the total product of gates generated equal to $K$. If the desired outcome is not achieved after these two additional applications of the protocol, then we continue to repeatedly apply the protocol in this way, until the desired outcome is finally achieved.

We can think of this as carrying out a random walk on the integers, where the position $x_n \in \mathbb{Z}$ after $n$ steps is the exponent of the product $K^{x_{n}}$ of gates after $n$ applications of the protocol. As such, the random walk starts from $x_{0} = 0$ and each step has an equal probability of moving the position to $x_{n+1} = x_{n}+1$ or $x_{n}-1$.

This random walk strategy for generating the $K$ gate by repeatedly converting $|K\rangle$ states into phase gates using TQF until successful is only viable if the probability of achieving the desired result goes to $1$ as the number $n$ of repeated applications of the protocol gets large ($n \rightarrow \infty$). We analyze the random walk in Appendix~\ref{sec:random_walk} and find that this is indeed the case, but that the probability of not achieving the desired result (a positive valued position) in $n$ steps, where $n$ is odd, is
\begin{equation}
\text{Prob}\left( x_1 , \ldots , x_{n} \leq 0 \right) = \frac{n !!}{(n+1)!!}
,
\end{equation}
which goes to zero as $\sqrt\frac{2}{\pi n}$ for $n$ large. This is certainly less ideal than an exponentially fast convergence (which is what we found for all our other probabilistic protocols), but it nonetheless permits a higher level strategy that allows the use of such random walk generated gates to be used in BQP (bounded error quantum polynomial time) quantum computations.

A strategy for using such random walk generated $K$ gates while satisfying the conditions for BQP is the following. If the quantum computation we wish to carry out involves $k$ applications of the gate $K$, then we allot each $K$ gate at most $k^2$ steps in the random walk used to generate it. For each $K$ gate, if the random walk reaches $k^2$ steps without achieving the desired result, we terminate the process and consider the gate and, hence, the entire computation to have failed. Each $K$ gate that we attempt to implement in this way will thus have a probability
\begin{equation}
p_{K\text{-fail}} = \frac{(k^2) !!}{(k^{2}+1)!!}
\end{equation}
of failing and $1- p_{K\text{-fail}}$ of being successfully generated. Consequently, the entire computation will have a probability $(1 - p_{K\text{-fail}} )^{k}$ of being successfully implemented (at least with respect to the $K$ gates' random walk issue). In the limit as $k$ gets large, we see that the probability for the entire computation being successfully implemented is
\begin{equation}
\lim_{k \rightarrow \infty} (1 - p_{K\text{-fail}})^{k} = \lim_{k \rightarrow \infty} \left( 1 - \sqrt{\frac{2}{\pi k^2}} \right)^{k} = e^{-\sqrt{\frac{2}{\pi}} }.
\end{equation}
Thus, the probability of failure of the computation due to the random walk generated $K$ gates using this implementation strategy is bounded, and hence this strategy satisfies the conditions for BQP. (We note that it can be inferred from the measurement outcomes whether or not the $K$ gate has been successfully applied, so there is no requirement for the success probability to be greater than $1/2$.) However, we notice that the number of operations that may be required to implement all $k$ applications of $K$ gates using this strategy scales as $k^3$, so the polynomial exponent of the computation length scaling increases by a factor of 3.

Clearly, the crude strategy described here is far from ideal, as our goal was merely to demonstrate the existence of a strategy that works, in principle, within BQP. There are a number of strategies for algorithm optimization. One simple way to reduce the negative impact (i.e. the costliness and convergence issues) of the irrational $K$ gate is to compile the quantum algorithm with an aversion to the $K$ gate. In other words, when synthesizing gates or algorithms from the generating gate set $\{ X, Z, B , K \}$, the optimization should minimize a weighted combination of both the word length and the total number of $K$ gates used.

Another strategy for improving the situation is to employ a sliding cutoff for the number of attempts allowed before terminating the random walk attempting to implement a $K$ gate. In particular, one may wish to terminate random walks sooner at earlier stages of the computation, while allowing longer walks as one nears the end of the computation. A more drastic modification of the strategy is to pick a cutoff point in each random walk (which could even be after one step), after which one recompiles the algorithm to see if a different path forward would be more economical~\cite{Bocharov2014}.

\section{Controlled-$Z$ Entangling gate}
\label{sec:CZ}

The set of single qubit gates $\{ X, Z, B , K \}$ is computationally universal for single qubits. If we can supplement this gate set with any entangling gate, we would obtain a full computationally universal gate set. In this section, we provide a protocol for generating the two qubit entangling gate, controlled-$Z$
\begin{equation}
C(Z) = \left[
\begin{array}{cccc}
1 & 0 & 0 & 0 \\
0 & 1 & 0 & 0 \\
0 & 0 & 1 & 0 \\
0 & 0 & 0 & -1
\end{array}
\right]
.
\end{equation}

In addition to the inexpensive gates and operations (such as topological charge measurements, braiding, and TQF), our protocol will utilize a single application of the Hadamard gate
\begin{equation}
H= \frac{1}{\sqrt{2}}
\left[
\begin{array}{cc}
1 & 1 \\
1 & -1
\end{array}
\right]
.
\end{equation}
The Hadamard gate is costly, as it requires a long string of the generating gates to accurately produce, including a large number of applications of the irrational phase gate $K$, each of which is relatively expensive to implement, as seen in the previous section. More specifically, Ref.~\cite{Bocharov-unpublished} developed algorithms for synthesizing single qubit gates from the set $\{ Z, B , K \}$ that produces approximate implementations of $H$ to precision $\varepsilon$ with $K$-count around $4 \log_{7} (1/ \varepsilon)$. For example, they produce an approximate $H$ gate with trace precision $\varepsilon < 10^{-8}$ using a string of 67 gates, 40 of which are $K$ gates~\cite{Bocharov-unpublished}.

However, the single application of $H$ is utilized during the ancillary state preparation stage of the protocol, so its generation can be carried out in a parallelized fashion that avoids the previously discussed issues with applying $K$ to the computational state within a quantum computation. The use of the Hadamard gate also means that the precision of our $C(Z)$ gate will depend upon the precision to which we approximate the Hadamard gate $H$ that is applied in this protocol.

We split the protocol for generating $C(Z)$ into two parts: (A) generating the ancillary two qubit states that will serve as entanglement resources, and (B) converting the entanglement resource ancillary state into the $C(Z)$ gate using TQF.

\subsection{Generating entanglement resources}

We wish to generate ancillary pairs of qubits in the 1221 encoding which are in the entangled two qubit state
\begin{equation}
| \Phi_{H} \rangle = H | \Phi^{+} \rangle = \frac{1}{2} \left( | 11 \rangle + | 13 \rangle + |31 \rangle - | 33 \rangle \right)
.
\end{equation}
[We use the conventional notation $| \Phi^{\pm} \rangle = \frac{1}{\sqrt{2}} ( | 11 \rangle \pm | 33 \rangle)$ and $| \Psi^{\pm} \rangle = \frac{1}{\sqrt{2}} ( | 13 \rangle \pm | 31 \rangle)$ for the Bell states.] The Hadamard operator here can be equivalently applied to either the first or the second qubit in this expression. We now provide a protocol for generating such states.

Starting from two ancillary qubits in the 1221 encoding (enumerating the quasiparticles 1-8 from left to right), we follow the steps:
\begin{enumerate}
\item Initialize each 1221 qubit in the state $|+\rangle  = H | 1 \rangle = \frac{1}{\sqrt{2}}(  | 1 \rangle + | 3 \rangle ) $.

\item Measure the collective topological charge $r \in \{ 0 , 2, 4 \}$ of quasiparticles 3, 4, 5, and 6. If $r\neq 0$, go to step 1. If $r=0$, go to step 3.

\item Perform a braiding operation that interchanges quasiparticles 3 and 4 and one that interchanges quasiparticles 5 and 6, with the opposite chirality.

\item Move the quasiparticles 3, 4, 5, and 6 to the right of quasiparticles 7 and 8 in a manner that does not create entanglement (i.e. have them all follow the same path).

\item Apply the Hadamard gate $H$ to one of the qubits.

\end{enumerate}

Step 1 may be carried out by applying a Hadamard gate $H$ to the qubit initialized in the basis state $|1\rangle$. However, this is not the best way to perform this initialization, since the Hadamard gate is costly and not exact. A better method is suggested by Eq.~(\ref{eq:1221_1122}) and the corresponding $F$-symbol
\begin{equation}
\left[ F^{122}_{1} \right]_{ab}
= \frac{1}{\sqrt{2}}
\left[
\begin{array}{cc}
1 & 1 \\
1 & -1
\end{array}
\right]_{ab}
= H_{ab}
\end{equation}
where $a=1,3$ and $b=0,2$. From this, we see that, for a qubit in the 1221 encoding, we can simply measure the topological charge $b \in \{0,2 \}$ of the pair of charge 2 quasiparticles. (We could, equivalently, measure the topological charge $b \in \{0,2 \}$ of the pair of charge 1 quasiparticles, if desired.) If the measurement outcome is $b=0$, then the 1221 qubit is in the desired state $|+\rangle$. If the measurement outcome is $b=2$, then the 1221 qubit is in the state $|-\rangle = Z |+\rangle$, so we merely apply the gate $Z$ to obtain the desired initial state. Similarly, we could initialize the 1221 qubit in the state $|+\rangle$ by pair producing a pair of charge 1 quasiparticles from vacuum and pair producing a pair of charge 2 quasiparticles from vacuum and then aligning them into the 1221 qubit configuration.

The measurement in Step 2 is diagrammatically given by
\begin{eqnarray}
&&
\pspicture[shift=*](-2.5,-1)(2.5,1)
\psset{unit=0.33cm}\pslabelsep=0.1cm\scriptsize
  \psline (-7,1)(-7,2)
  \psline (-5,1)(-5,2)
  \psline (-3,1)(-3,2)
  \psline (-1,1)(-1,2)
  \psline (1,1)(1,2)
  \psline (3,1)(3,2)
  \psline (5,1)(5,2)
  \psline (7,1)(7,2)
  \psarc (-6,1){1}{180}{0}
  \psarc (-2,1){1}{180}{0}
  \psarc (2,1){1}{180}{0}
  \psarc (6,1){1}{180}{0}
  \psarc (-4,0){2}{180}{0}
  \psarc (4,0){2}{180}{0}
  \uput[90] (-7,2){$1$}
  \uput[90] (-5,2){$2$}
  \uput[90] (-3,2){$2$}
  \uput[90] (-1,2){$1$}
  \uput[90] (1,2){$1$}
  \uput[90] (3,2){$2$}
  \uput[90] (5,2){$2$}
  \uput[90] (7,2){$1$}
  \uput[-90] (-4,-2){$a$}
  \uput[-90] (4,-2){$b$}
\endpspicture
\notag \\
&& \mapsto
\pspicture[shift=*](-2.5,-1)(2.5,1.2)
\psset{unit=0.33cm}\pslabelsep=0.1cm\scriptsize
  \psline (-7,1)(-7,2)
  \psline (-5,1)(-5,2)
  \psline (5,1)(5,2)
  \psline (7,1)(7,2)
  \psarc (-6,1){1}{180}{0}
  \psarc (-2,1){1}{180}{0}
  \psarc (2,1){1}{180}{0}
  \psarc (6,1){1}{180}{0}
  \psarc (-3.125,1){0.25}{90}{-90}
  \psarc (3.125,1){0.25}{-90}{90}
  \psline (-3.125,1.25)(3.125,1.25)
  \psline[border=2pt] (-3.125,0.75)(3.125,0.75)
  \psline[border=2pt] (-3,1)(-3,2)
  \psline[border=2pt] (-1,1)(-1,2)
  \psline[border=2pt] (1,1)(1,2)
  \psline[border=2pt] (3,1)(3,2)
  \psarc (-4,0){2}{180}{0}
  \psarc (4,0){2}{180}{0}
  \uput[90] (-7,2){$1$}
  \uput[90] (-5,2){$2$}
  \uput[90] (-3,2){$2$}
  \uput[90] (-1,2){$1$}
  \uput[90] (1,2){$1$}
  \uput[90] (3,2){$2$}
  \uput[90] (5,2){$2$}
  \uput[90] (7,2){$1$}
  \uput[-90] (0,0.75){$w_r$}
  \uput[-90] (-4,-2){$a$}
  \uput[-90] (4,-2){$b$}
\endpspicture
\end{eqnarray}
When $r=0$, this is equal to
\begin{equation}
\delta_{ab} \frac{1}{d_{a}}
\pspicture[shift=*](-2.5,-1.5)(2.5,1.0)
\psset{unit=0.33cm}\pslabelsep=0.1cm\scriptsize
  \psellipse (0,0)(6,4)
  \psframe*[linecolor=white] (-6,4.1)(6,0)
  \psarc (-6,1){1}{180}{0}
  \psarc (-2,1){1}{180}{0}
  \psarc (2,1){1}{180}{0}
  \psarc (6,1){1}{180}{0}
  \psarc (0,0){2}{180}{0}
  \uput[90] (-7,1){$1$}
  \uput[90] (-5,1){$2$}
  \uput[90] (-3,1){$2$}
  \uput[90] (-1,1){$1$}
  \uput[90] (1,1){$1$}
  \uput[90] (3,1){$2$}
  \uput[90] (5,1){$2$}
  \uput[90] (7,1){$1$}
  \uput[-90] (0,-2){$a$}
  \uput[-90] (0,-4){$a$}
\endpspicture
\end{equation}

Steps 3 and 4 are simply rearranging the configuration of quasiparticles to that of two qubits in the 1221 encoding, in a manner that does not affect the state. Now, the original quasiparticles 1, 2, 7, and 8 comprise the first 1221 qubit and quasiparticles 4, 3, 6, and 5 comprise the second 1221 qubit.

Combining steps 2, 3, and 4 (with measurement outcome $r=0$), the effect on basis states can be written diagrammatically as
\begin{eqnarray}
&&
\pspicture[shift=*](-2.5,-1)(2.5,1)
\psset{unit=0.33cm}\pslabelsep=0.1cm\scriptsize
  \psline (-7,1)(-7,2)
  \psline (-5,1)(-5,2)
  \psline (-3,1)(-3,2)
  \psline (-1,1)(-1,2)
  \psline (1,1)(1,2)
  \psline (3,1)(3,2)
  \psline (5,1)(5,2)
  \psline (7,1)(7,2)
  \psarc (-6,1){1}{180}{0}
  \psarc (-2,1){1}{180}{0}
  \psarc (2,1){1}{180}{0}
  \psarc (6,1){1}{180}{0}
  \psarc (-4,0){2}{180}{0}
  \psarc (4,0){2}{180}{0}
  \uput[90] (-7,2){$1$}
  \uput[90] (-5,2){$2$}
  \uput[90] (-3,2){$2$}
  \uput[90] (-1,2){$1$}
  \uput[90] (1,2){$1$}
  \uput[90] (3,2){$2$}
  \uput[90] (5,2){$2$}
  \uput[90] (7,2){$1$}
  \uput[-90] (-4,-2){$a$}
  \uput[-90] (4,-2){$b$}
\endpspicture
\notag \\
&& \mapsto \delta_{ab} \frac{1}{\sqrt{3}}
\pspicture[shift=*](-2.5,-1)(2.5,1)
\psset{unit=0.33cm}\pslabelsep=0.1cm\scriptsize
  \psline (-7,1)(-7,2)
  \psline (-5,1)(-5,2)
  \psline (-3,1)(-3,2)
  \psline (-1,1)(-1,2)
  \psline (1,1)(1,2)
  \psline (3,1)(3,2)
  \psline (5,1)(5,2)
  \psline (7,1)(7,2)
  \psarc (-6,1){1}{180}{0}
  \psarc (-2,1){1}{180}{0}
  \psarc (2,1){1}{180}{0}
  \psarc (6,1){1}{180}{0}
  \psarc (-4,0){2}{180}{0}
  \psarc (4,0){2}{180}{0}
  \uput[90] (-7,2){$1$}
  \uput[90] (-5,2){$2$}
  \uput[90] (-3,2){$2$}
  \uput[90] (-1,2){$1$}
  \uput[90] (1,2){$1$}
  \uput[90] (3,2){$2$}
  \uput[90] (5,2){$2$}
  \uput[90] (7,2){$1$}
  \uput[-90] (-4,-2){$a$}
  \uput[-90] (4,-2){$a$}
\endpspicture
.
\end{eqnarray}
The effect on the initialized states (i.e. combining steps 1-4) is thus
\begin{equation}
|+\rangle |+\rangle \mapsto |\Phi^{+} \rangle
.
\end{equation}

Step 5 takes us from the Bell state $|\Phi^{+} \rangle$, obtained by applying steps 1-4, to the desired state
\begin{equation}
|\Phi_{H} \rangle = \openone \otimes H |\Phi^{+} \rangle = H \otimes \openone |\Phi^{+} \rangle
.
\end{equation}
In contrast with the initialization step 1, we do not know a way to circumvent the use of the costly Hadamard gate $H$ in this last step.

\subsection{Converting $|\Phi_{H} \rangle$ into a controlled-$Z$ gate}

Assuming that we now have two ancillary qubits in the 1221 encoding paired up in the entangled two qubit state $|\Phi_{H} \rangle$, we can convert this ancillary state into the $C(Z)$ gate acting on two computational qubits in the 1221 encoding using the following protocol.

Starting from two computational qubits ($A$ and $B$) in the 1221 encodings and an ancillary pair of qubits (1 and 2) in the 1221 encoding in the entangled state $|\Phi_{H} \rangle$, we follow the steps:

\begin{enumerate}
\item Perform TQF on qubits $A$ and $1$, with measurement outcome $z_{A1} \in \{0,4\}$. If $z_{A1} =4$, apply the gate $Z$ to qubit $A$.

\item Perform TQF on qubits $B$ and $2$, with measurement outcome $z_{B2} \in \{0,4\}$. If $z_{B2} =4$, apply the gate $Z$ to qubit $B$.

\end{enumerate}

It is straightforward to check that the effect of this protocol, which can be written for the computational qubit basis states as
\begin{eqnarray}
|a,b \rangle |\Phi_{H}\rangle &\mapsto& 2 Q^{(0)}_{A1} Q^{(0)}_{B2} |a,b \rangle |\Phi_{H}\rangle  \\
|a,b \rangle |\Phi_{H}\rangle &\mapsto& 2 Z_{B} Q^{(0)}_{A1} Q^{(4)}_{B2} |a,b \rangle |\Phi_{H}\rangle  \\
|a,b \rangle |\Phi_{H}\rangle &\mapsto& 2 Z_{A} Q^{(4)}_{A1} Q^{(0)}_{B2} |a,b \rangle |\Phi_{H}\rangle  \\
|a,b \rangle |\Phi_{H}\rangle &\mapsto& 2 Z_{A} Z_{B} Q^{(4)}_{A1} Q^{(4)}_{B2} |a,b \rangle |\Phi_{H}\rangle
\end{eqnarray}
for the four possible combinations of TQF measurement outcomes $z_{A1}$ and $z_{B2}$, all give the same result, which is an application of the controlled-$Z$ gate to the two computational qubits
\begin{equation}
|a,b \rangle |\Phi_{H} \rangle \mapsto C(Z) |a,b \rangle
.
\end{equation}
We emphasize that this protocol consumes the ancillary entanglement resource pair of qubits in order to convert them into the $C(Z)$ gate acting on qubits $A$ and $B$ of the computational state.

\section{Conclusion}
\label{sec:discussion}

In the example of SU$(2)_4$ and JK$_{4}$ anyons, we have seen that it is possible for an anyon model to not be computationally universal with braiding alone, but to become computationally universal when braiding is supplemented with fusion and measurement operations. This demonstration of the utility of fusion and measurement operations encourages further analysis of such operations and exploration for additional theories whose computational power can be supplemented or even made computationally universal in this way, particularly for theories that would be easier to physically realize.

The same strategy of supplementing braiding operations by fusion and measurement operations can also be considered for theories with symmetry defects, which are described by $G$-crossed braided categories, in which braiding is generalized to incorporate symmetry action~\cite{Barkeshli14}. The defect theories that are relatively easy to physically realize appear to have ($G$-crossed) braiding which is not computationally universal, so it would be interesting to determine whether any of them could be made computationally universal via fusion and measurement operations. One known example of a defect theory which may benefit from fusion and measurement operations is a bilayer Ising TQFT system with layer interchange symmetry ($G=\mathbb{Z}_2$); fusion and measurement may be used to switch between a qubit state being encoded in $\sigma$ quasiparticles, where they may be relatively easier to physically manipulate, and being encoded in defects (also known as ``genons''), which provide computational universality through braiding~\cite{Barkeshli2013}.

\appendix

\section{Basic Data of the Anyon Models}
\label{sec:basic_data}

In this section, we provide general expressions for the basic data (topological charges, fusion rules, $F$-symbols, and $R$-symbols) of the SU$(2)_k$ and JK$_k$ anyon models, for general $k$. For JK$_4$, we also evaluate and tabulate the $F$-symbols and $R$-symbols that we use in this paper, for convenience.

\subsection{$\text{SU}\left(2\right)_{k}$ Anyons}
\label{sec:SU(2)_k}

The SU$\left( 2\right) _{k}$ anyon models (for $k$ an integer) are ``$q$-deformed'' versions of the usual SU$\left( 2\right) $ representation theory~\cite{Chari-book}. Roughly speaking, this means integers $n$ are replaced by ``$q$-integers'' $\left[ n\right]_{q}\equiv \frac{q^{n/2}-q^{-n/2}}{q^{1/2}-q^{-1/2}}$, where the deformation parameter $q=e^{i\frac{2\pi }{k+2}}$ is taken to be a simple root of unity. These anyon models describe SU$\left(2\right) _{k}$ Chern-Simons theories~\cite{Witten89} and WZW CFTs~\cite{Wess71,Witten83}, and give rise to the Jones polynomials of knot theory~\cite{Jones85}. Their braiding statistics are known~\cite{Freedman02b} to be computationally universal for all $k$, except $k=1$, $2$, and $4$. The anyon models are described by:
\begin{widetext}
\begin{equation*}
\begin{tabular}{|l|l|}
\hline
\multicolumn{2}{|l|}{$\mathcal{C}=\left\{ 0,\frac{1}{2},\ldots ,\frac{k}{2}\right\}, \quad j_{1}\times j_{2}=\sum\limits_{j=\left|
j_{1}-j_{2}\right| }^{\min \left\{ j_{1}+j_{2},k-j_{1}-j_{2}\right\} }j$} \\
\hline
\multicolumn{2}{|l|}{$\left[ F_{j}^{j_{1},j_{2},j_{3}}\right]
_{j_{12},j_{23}}=\left( -1\right) ^{j_{1}+j_{2}+j_{3}+j}\sqrt{\left[
2j_{12}+1\right] _{q}\left[ 2j_{23}+1\right] _{q}}\left\{
\begin{array}{ccc}
j_{1} & j_{2} & j_{12} \\
j_{3} & j & j_{23}%
\end{array}%
\right\} _{q}^{\phantom{T}},$} \\
\multicolumn{2}{|l|}{$\left\{
\begin{array}{ccc}
j_{1} & j_{2} & j_{12} \\
j_{3} & j & j_{23}%
\end{array}%
\right\} _{q}=\Delta \left( j_{1},j_{2},j_{12}\right) \Delta \left(
j_{12},j_{3},j\right) \Delta \left( j_{2},j_{3},j_{23}\right) \Delta \left(
j_{1},j_{23},j\right) $} \\
\multicolumn{2}{|l|}{$\quad \quad \quad \quad \quad \quad \quad \quad \times
\sum\limits_{z}\left\{ \frac{\left( -1\right) ^{z}\left[ z+1\right] _{q}!}{%
\left[ z-j_{1}-j_{2}-j_{12}\right] _{q}!\left[ z-j_{12}-j_{3}-j\right] _{q}!%
\left[ z-j_{2}-j_{3}-j_{23}\right] _{q}!\left[ z-j_{1}-j_{23}-j\right] _{q}!}%
\right. $} \\
\multicolumn{2}{|l|}{$\quad \quad \quad \quad \quad \quad \quad \quad \quad
\quad \quad \times \left. \frac{1}{\left[ j_{1}+j_{2}+j_{3}+j-z\right] _{q}!%
\left[ j_{1}+j_{12}+j_{3}+j_{23}-z\right] _{q}!\left[ j_{2}+j_{12}+j+j_{23}-z%
\right] _{q}!}\right\}_{\phantom{g}} ,$} \\
\multicolumn{2}{|l|}{$\Delta \left( j_{1},j_{2},j_{3}\right) =\sqrt{\frac{%
\left[ -j_{1}+j_{2}+j_{3}\right] _{q}!\left[ j_{1}-j_{2}+j_{3}\right] _{q}!%
\left[ j_{1}+j_{2}-j_{3}\right] _{q}!}{\left[ j_{1}+j_{2}+j_{3}+1\right]
_{q}!}}^{\phantom{T}}_{\phantom{g}},\quad \quad \quad \left[ n\right] _{q}! \equiv \prod\limits_{m=1}^{n}%
\left[ m\right] _{q}$} \\ \hline
\multicolumn{2}{|l|}{$R_{j}^{j_{1},j_{2}}=\left( -1\right)
^{j-j_{1}-j_{2}}q^{\frac{1}{2}\left( j\left( j+1\right) -j_{1}\left(
j_{1}+1\right) -j_{2}\left( j_{2}+1\right) \right) }$} \\ \hline
$d_{j}= \left[ 2j+1 \right]_{q} =\frac{\sin \left( \frac{\left( 2j+1\right) \pi }{k+2}\right)^{\phantom{T}} }{\sin
\left( \frac{\pi }{k+2}\right)_{\phantom{g}} }, \quad \mathcal{D}=\frac{\sqrt{ \frac{k+2}{2}}}{\sin \left( \frac{\pi}{k+2} \right)_{\phantom{g}}}$ &
$\varkappa_j = \left(-1\right)^{2j}$ \\ \hline
$\theta _{j} = q^{j(j+1)} =e^{i2\pi \frac{j\left(
j+1\right) }{k+2}}$ & $S_{j_1 j_2} = \sqrt{\frac{2}{k+2}} \sin \left( \frac{\left(2j_1 +1 \right)\left(2j_2 +1 \right) \pi}{k+2} \right)$ \\ \hline
\end{tabular}%
\end{equation*}
where $\left\{ \quad \right\} _{q}$ is a ``$q$-deformed'' version of the usual SU$\left( 2\right) $ $6j$-symbols. The sum over $z$ in this expression is over all integers for which each term is well-defined, i.e. $z_{\text{min}} \leq z \leq z_{\text{max}}$, where $z_{\text{min}} = \max \{ j_{1}+j_{2}+j_{3}+j , j_{1}+j_{12}+j_{3}+j_{23},  j_{2}+j_{12}+j+j_{23} \}$ and $z_{\text{max}} = \min \{ j_{1}+j_{2}+j_{12}, j_{12}+j_{3}+j, j_{2}+j_{3}+j_{23}, j_{1}+j_{23}+j \}$. The Frobenius-Schur indicator $\varkappa_{j} = d_{j} \left[ F^{jjj}_{j} \right]_{0 0}$ is a gauge invariant quantity which plays a role in bending and straightening lines.

\subsection{JK$_k$ Anyons}
\label{sec:JK_k}

There are anyon models based on the Jones-Kauffman bracket~\cite{Kauffman-book,Wang-book}, which may be derived from Temperley-Lieb recoupling theory. These anyon models, which we denote as JK$_k$, are closely related to the SU$\left( 2\right) _{k}$ anyon models. They have the same number of topological charges and the same fusion rules as SU$\left( 2\right) _{k}$ at the corresponding level, and the basic data shares a very similar structure. It is conventional to label the topological charges of JK$_k$ by integers, rather than both integers and half-integers, so there will be a translation between these by multiplying/dividing the charge labels by 2. We define $A=ie^{-i\frac{\pi }{2(k+2)}}$, and $\left[ n\right]_{A}\equiv \frac{A^{2n}-A^{-2n}}{A^{2}-A^{-2}}$. The anyon model data is given by:%
\begin{equation*}
\begin{tabular}{|l|l|}
\hline
\multicolumn{2}{|l|}{$\mathcal{C}=\left\{ 0,1,\ldots ,k\right\}, \quad a\times b= \left|
a-b\right| + \left( \left|a-b\right| +2 \right)  + \ldots + \min \left\{ a+b,2k-a-b\right\} $ } \\
\hline
\multicolumn{2}{|l|}{$\left[ F_{d}^{abc}\right]_{ef}= \frac{ \sqrt{d_{e} d_{f}} }{\sqrt{ \theta(a,b,e) \theta(c,d,e) \theta(b,c,f) \theta(a,d,f) }} \text{Tet} \left[
\begin{array}{ccc}
a & b & e \\
c & d & f%
\end{array}%
\right]_{A}^{\phantom{T}},$} \\
\multicolumn{2}{|l|}{$\text{Tet}\left[
\begin{array}{ccc}
a & b & e \\
c & d & f%
\end{array}%
\right] _{A}= \frac{ \mathcal{I}! }{ \mathcal{E}!} \sum\limits_{z}\left\{ \frac{\left( -1\right) ^{z}\left[ z+1\right] _{A}!}{%
\left[ z-\frac{a+b+e}{2}\right] _{A}!\left[ z-\frac{e+c+d}{2}\right] _{A}!%
\left[ z-\frac{b+c+f}{2}\right] _{A}!\left[ z-\frac{a+f+d}{2}\right] _{A}!}%
\right. $} \\
\multicolumn{2}{|l|}{$\quad \quad \quad \quad \quad \quad \quad \quad \quad \quad \quad
\times \left. \frac{1}{\left[ \frac{a+b+c+d}{2}-z\right] _{A}! \left[ \frac{a+e+c+f}{2}-z\right] _{A}!\left[ \frac{b+e+d+f}{2} -z \right] _{A}!}\right\}_{\phantom{g}} ,$} \\
\multicolumn{2}{|l|}{$\theta \left( a,b,c\right) = \frac{\left[ \frac{a+b+c}{2}+1\right]_{A}! \left[ \frac{-a+b+c}{2} \right] _{A}!\left[ \frac{a-b+c}{2}\right] _{A}! \left[ \frac{a+b-c}{2}\right] _{A}!}{\left[ a\right]_{A}! \left[ b\right]_{A}!  \left[ c\right]_{A}! }^{\phantom{T}}_{\phantom{g}},
\quad \quad \quad \left[ n\right] _{A}! \equiv \prod\limits_{m=1}^{n} \left[ m\right] _{A}$} \\
\multicolumn{2}{|l|}{$\left. \mathcal{E}! = \left[ a \right]_{A}! \left[ b \right]_{A}! \left[ c \right]_{A}! \left[ d \right]_{A}! \left[ e \right]_{A}! \left[ f \right]_{A}! \right. ^{\phantom{T}}_{\phantom{g}},$} \\
\multicolumn{2}{|l|}{$\left. \mathcal{I}! = \left[ \frac{-a+b+e}{2} \right] _{A}!\left[ \frac{a-b+e}{2}\right] _{A}! \left[ \frac{a+b-e}{2}\right] _{A}!\left[ \frac{-c+d+e}{2} \right] _{A}!\left[ \frac{c-d+e}{2}\right] _{A}! \left[ \frac{c+d-e}{2}\right] _{A}! \right.^{\phantom{T}}_{\phantom{g}} $}\\
\multicolumn{2}{|l|}{$\qquad \quad \left. \times \left[ \frac{-a+d+f}{2} \right] _{A}!\left[ \frac{a-d+f}{2}\right] _{A}! \left[ \frac{a+d-f}{2}\right] _{A}! \left[ \frac{-b+c+f}{2} \right] _{A}!\left[ \frac{b-c+f}{2}\right] _{A}! \left[ \frac{b+c-f}{2}\right] _{A}! \right.^{\phantom{T}}_{\phantom{g}} $}\\
\hline
\multicolumn{2}{|l|}{$R_{c}^{ab}=\left( -1\right)
^{\frac{a+b-c}{2}}A^{\frac{1}{2}\left( c\left( c+2\right) -a\left(
a+2\right) -b\left( b+2\right) \right) }$} \\ \hline
$d_{a}= \left(-1\right)^{a}\left[ a+1 \right]_{A} =\frac{\sin \left( \frac{\left( a+1\right) \pi }{k+2}\right)^{\phantom{T}} }{\sin
\left( \frac{\pi }{k+2}\right)_{\phantom{g}} }, \quad \mathcal{D}=\frac{\sqrt{ \frac{k+2}{2}}}{\sin \left( \frac{\pi}{k+2} \right)_{\phantom{g}}}$ & $\varkappa _{a}=1$ \\ \hline
$\theta _{a}=(-1)^{a} A^{a(a+2)} = i^{- a^{2}} e^{-i 2 \pi \frac{a (a+2)}{ 4 (k+2)}} $  &  $S_{ab}=\sqrt{\frac{2}{k+2}} (-1)^{ab} \sin \left( \frac{\left(
a+1\right) \left( b+1\right) \pi }{k+2}\right) $ \\ \hline
\end{tabular}%
\end{equation*}
\end{widetext}
The sum over $z$ in this expression is over all integers for which each term is well-defined, i.e. $z_{\text{min}} \leq z \leq z_{\text{max}}$, where $z_{\text{min}} = \max \{ a+b+c+d , a+e+c+f,  b+e+d+f \}$ and $z_{\text{max}} = \min \{ a+b+e, e+c+d, b+c+f, a+f+d \}$. It is straightforward to check that the conditions for complete isotopy invariance are satisfied for JK$_{k}$, i.e. all the bending transformations in this theory (for the choice of gauge used above) are trivial, since $\left[ F^{aab}_{b} \right]_{0 c} = \left[ F^{abb}_{a} \right]_{c 0} = \sqrt{ \frac{d_c}{ d_a d_b } } $ for all $a,b,c$ with $N_{ab}^{c}\neq 0$. We note that $k=2$ is the Ising anyon model.

Comparing the basic data of JK$_k$ to that of SU$(2)_k$, we notice that the structure of these expressions and their SU$(2)_k$ counterparts are nearly identical. The main difference is that the $q$-deformation factor is changed by a minus sign (note that $A^2$ plays the role of $q$), which requires minus sign factors that necessarily arise in the associativity expressions, as well as factors of $i$ that arise in the braiding. Notice that the Frobenius-Schur indicator $\varkappa_{a} = d_{a} \left[ F^{aaa}_{a} \right]_{0 0}$ here is trivial, as required for the theory to be completely isotopy invariant.

In fact, the two theories can be related by gluing a semion onto the odd/half-integer charges. More specifically, this means we take the direct product of one of these theories with a semion and then restrict the topological charge set so that odd charges in the JK$_k$ sector (or half-integer charges in the SU$(2)_k$ sector) and the nontrivial charge in the semion sector are always paired up. We can write this as $\text{SU}\left( 2\right) _{k} = \left. \overline{\text{JK}_{k} } \times \mathbb{Z}_2^{(3/2)} \right|_{\mathcal{C}}$, where $\mathbb{Z}_2^{(3/2)} = \overline{ \text{SU}\left( 2\right) _{1} }$ is the ($\mathbb{Z}_2$) semion with $F^{111}_{1} = -1$ and $\theta_1 = -i$, and $\mathcal{C} = \{ (a,b) \, | \,  a \in \{ 0 , 1 , \ldots , k \}, b \in \{0,1\}, a \equiv b \mod 2 \}$. This relation is confirmed by the corresponding values of the topological twists and $S$-matrices of the respective theories.

\subsection{Some Useful $F$-symbols and $R$-symbols of JK$_4$}
  \label{sec:JK_4}

In this section, we tabulate the $F$-symbols and $R$-symbols of JK$_4$ that are used for calculations in this paper, for convenience.
\begin{equation*}
\begin{tabular}{|l|}
\hline
$\left[ F_{d}^{abc}\right]_{ef}= 0$ if $N_{ab}^{e}N_{ec}^{d} = 0$ or $N_{bc}^{f} N_{af}^{d} = 0$
\\
$\left[ F_{d}^{abc}\right]_{ef}= 1$ if $a$, $b$, $c$, or $d=0$ and $N_{ab}^{e}N_{ec}^{d} = N_{bc}^{f} N_{af}^{d} \neq 0$
\\
$\left[ F_{b}^{aab}\right]_{0c}= \left[ F_{a}^{abb}\right]_{c0} = \sqrt{\frac{d_c }{d_a d_b}}$ if $N_{ab}^{c} \neq 0$
\\
$\left[ F_{1}^{122}\right]_{10}= \left[ F_{1}^{122}\right]_{12}= \left[ F_{1}^{122}\right]_{30}= \frac{1}{\sqrt{2}} , \left[ F_{1}^{122}\right]_{32}= -\frac{1}{\sqrt{2}} $
\\
$\left[ F_{0}^{121}\right]_{11}= 1, \left[ F_{2}^{121}\right]_{11}= - \frac{1}{2} , \left[ F_{2}^{121}\right]_{13}= \frac{\sqrt{3}}{2}$
\\
$\left[ F_{1}^{111}\right]_{02}= \sqrt{\frac{2}{3}}, \left[ F_{1}^{111}\right]_{22}= -\frac{1}{\sqrt{3}}, \left[ F_{3}^{111}\right]_{22}= 1$
\\
$\left[ F_{0}^{321}\right]_{13}=1, \left[ F_{2}^{321}\right]_{11}= \frac{\sqrt{3}}{2}, \left[ F_{2}^{321}\right]_{13}= \frac{1}{2} $
\\
$\left[ F_{3}^{113}\right]_{02}= \sqrt{\frac{2}{3}}, \left[ F_{1}^{113}\right]_{22}= 1, \left[ F_{3}^{113}\right]_{22}= \frac{1}{\sqrt{3}}$
\\
$\left[ F_{1}^{111}\right]_{00}= \frac{1}{\sqrt{3}}, \left[ F_{1}^{111}\right]_{20}= \sqrt{\frac{2}{3}}$
\\
$\left[ F_{3}^{122}\right]_{14}= \left[ F_{1}^{322}\right]_{14}= \left[ F_{3}^{122}\right]_{34}= \left[ F_{1}^{322}\right]_{34}=  \frac{1}{\sqrt{2}}$
\\
$\left[ F_{3}^{421}\right]_{21}= \left[ F_{1}^{421}\right]_{23}= 1$
\\
\hline
$R^{11}_{0} = e^{-i \frac{\pi}{4}} , R^{11}_{2} = e^{i \frac{5\pi}{12}} $
\\
$R^{12}_{1} = R^{21}_{1} = e^{-i \frac{2 \pi}{3}},  R^{12}_{3} = R^{21}_{3} = e^{i \frac{5 \pi}{6}}$
\\
$R^{22}_{0} = e^{i \frac{2 \pi}{3}}, R^{22}_{2} = e^{-i \frac{2 \pi}{3}} , R^{22}_{4} = e^{-i \frac{\pi}{3}}$
\\
\hline
\end{tabular}%
\end{equation*}

\section{Random Walk}
\label{sec:random_walk}

In this section, we consider a random walk over the integers $\mathbb{Z}$, starting at position $x=0$, with each step having equal probability $\frac{1}{2}$ of taking a step from $x$ to $x+1$ or $x-1$. We compute the probability of never taking a positive valued position during a walk of $n$ steps or, equivalently, of never taking a negative valued position during a walk of $n$ steps. Since a walk starting at 0 can only become positive on an odd step, it is clear that this probability is the same for $n$ and $n+1$ steps, when $n$ is odd. As the problem is symmetric between avoiding positive and negative positions, we will simplify notation by computing the probability of the position never going negative within $n$ steps. We compute this iteratively by counting the total number of paths one can take without ever going negative within $n$ steps for $n=2m-1$ odd.

First we note that the total number of unconstrained paths possible after $n=2m-1$ steps is $N_{\text{total}}^{(m)} = 2^{n} = 2^{2m-1}$.

After the first step ($m=1$), there is equal probability of going to $x=+1$ and $-1$, so the number of paths that end up at positive position $x$ without ever going negative is simply
\begin{equation}
N_{x}^{(1)}= \delta_{x,1}.
\end{equation}

Since each step has equal probability of moving the position by $+1$ or $-1$, a consecutive pair of steps can move the position by $+2$, $0$, or $-2$, with there being one way to move $\pm 2$ and two ways to remain in the same position (i.e. first step forward and then step backward, or first step backward and then step forward). Thus, we can write the iterative expression for the number $N_{x}^{(m)}$ of possible $n=2m-1$ step paths that end up at positive position $x$ without ever going negative as
\begin{equation}
\left\{
\begin{array}{l l l}
N_{1}^{(m+1)} =  2 N_{1}^{(m)} + N_{3}^{(m)} & & \text{for} \quad x=1 \\
N_{x}^{(m+1)} = N_{x-2}^{(m)} + 2 N_{x}^{(m)} + N_{x+2}^{(m)} & & \text{for} \quad x>1
\end{array}
\right.
.
\end{equation}
(We notice that the contribution from position $x-2$ is missing from $x=1$, because that would have included a path that went negative.)

The total number of possible $n=2m-1$ step paths that end up at positive position without ever going negative is thus
\begin{equation}
N_{+}^{(m)} = \sum_{x>0} N_{x}^{(m)}
,
\end{equation}
which satisfies the iterative expression
\begin{equation}
N_{+}^{(m+1)} =  4 N_{+}^{(m)} - N_{1}^{(m)}
.
\end{equation}
Thus, the probability
\begin{equation}
p_{+}^{(m)} = \frac{N_{+}^{(m)} }{N_{\text{total}}^{(m)}}
\end{equation}
of never going negative within $n=2m-1$ steps can similarly be expressed iteratively as
\begin{eqnarray}
p_{+}^{(m+1)} &=& \frac{N_{+}^{(m+1)} }{N_{\text{total}}^{(m+1)}} = \frac{ 4 N_{+}^{(m)} - N_{1}^{(m)} }{ 4 N_{\text{total}}^{(m)}  } \notag \\
&=& \left( 1 - \frac{ N_{1}^{(m)} }{ 4 N_{+}^{(m)} } \right) p_{+}^{(m)} \notag \\
&=& \left( 1 - \frac{ 1}{ 2(m+1) } \right) p_{+}^{(m)}
,
\label{eq:p_it}
\end{eqnarray}
where we used the property
\begin{equation}
 N_{1}^{(m)} = \frac{ 2}{ m+1 } N_{+}^{(m)}
.
\label{eq:N_1_N_+}
\end{equation}
Solving the resulting iterative expression of Eq.~(\ref{eq:p_it}), we obtain the probability of avoiding going negative in $n=2m-1$ steps to be
\begin{equation}
p_{+}^{(m)} = \prod_{p=1}^{m} \left( 1 - \frac{1}{2p} \right) = \frac{ (2m-1)!!}{ (2m)!! }
.
\end{equation}

In order to obtain the scaling of $p_{+}^{(m)}$ for $m$ large, we use the relations of double factorials to the Gamma function together with Stirling's formula
\begin{eqnarray}
p_{+}^{(m)} &=& \frac{ \Gamma (m+\frac{1}{2})}{ \sqrt{\pi} \Gamma(m+1) } \notag \\
&\sim & \frac{1}{\sqrt{\pi}} \exp \left[ m \ln (m+\frac{1}{2}) -(m+\frac{1}{2}) + \ldots \right.
\notag \\
&& \left. \qquad - (m+\frac{1}{2}) \ln (m+1) + (m+1) +\ldots   \right]
\notag \\
&=&  \frac{1}{\sqrt{\pi}} m^{-\frac{1}{2}} + O(m^{-\frac{3}{2}})
.
\end{eqnarray}
Thus, we find that the probability of taking $n$ steps without ever taking a negative valued position goes to zero as $n^{-\frac{1}{2}}$ as $n \rightarrow \infty$.


\begin{thebibliography}{40}
\expandafter\ifx\csname natexlab\endcsname\relax\def\natexlab#1{#1}\fi
\expandafter\ifx\csname bibnamefont\endcsname\relax
  \def\bibnamefont#1{#1}\fi
\expandafter\ifx\csname bibfnamefont\endcsname\relax
  \def\bibfnamefont#1{#1}\fi
\expandafter\ifx\csname citenamefont\endcsname\relax
  \def\citenamefont#1{#1}\fi
\expandafter\ifx\csname url\endcsname\relax
  \def\url#1{\texttt{#1}}\fi
\expandafter\ifx\csname urlprefix\endcsname\relax\def\urlprefix{URL }\fi
\providecommand{\bibinfo}[2]{#2}
\providecommand{\eprint}[2][]{\url{#2}}

\bibitem[{\citenamefont{Kitaev}(2003)}]{Kitaev03}
\bibinfo{author}{\bibfnamefont{A.~Y.} \bibnamefont{Kitaev}},
  \bibinfo{journal}{Annals Phys.} \textbf{\bibinfo{volume}{303}},
  \bibinfo{pages}{2} (\bibinfo{year}{2003}), \eprint{quant-ph/9707021}.

\bibitem[{\citenamefont{Freedman}(1998)}]{Freedman98}
\bibinfo{author}{\bibfnamefont{M.~H.} \bibnamefont{Freedman}},
  \bibinfo{journal}{Proc. Natl. Acad. Sci. USA} \textbf{\bibinfo{volume}{95}},
  \bibinfo{pages}{98} (\bibinfo{year}{1998}).

\bibitem[{\citenamefont{Preskill}(1998)}]{Preskill98}
\bibinfo{author}{\bibfnamefont{J.}~\bibnamefont{Preskill}}, in
  \emph{\bibinfo{booktitle}{Introduction to Quantum Computation}}, edited by
  \bibinfo{editor}{\bibfnamefont{H.-K.} \bibnamefont{Lo}},
  \bibinfo{editor}{\bibfnamefont{S.}~\bibnamefont{Popescu}}, \bibnamefont{and}
  \bibinfo{editor}{\bibfnamefont{T.~P.} \bibnamefont{Spiller}}
  (\bibinfo{publisher}{World Scientific}, \bibinfo{year}{1998}),
  \eprint{quant-ph/9712048}.

\bibitem[{\citenamefont{Freedman
  et~al.}(2002{\natexlab{a}})\citenamefont{Freedman, Larsen, and
  Wang}}]{Freedman02a}
\bibinfo{author}{\bibfnamefont{M.~H.} \bibnamefont{Freedman}},
  \bibinfo{author}{\bibfnamefont{M.~J.} \bibnamefont{Larsen}},
  \bibnamefont{and} \bibinfo{author}{\bibfnamefont{Z.}~\bibnamefont{Wang}},
  \bibinfo{journal}{Commun. Math. Phys.} \textbf{\bibinfo{volume}{227}},
  \bibinfo{pages}{605} (\bibinfo{year}{2002}{\natexlab{a}}),
  \eprint{quant-ph/0001108}.

\bibitem[{\citenamefont{Freedman
  et~al.}(2002{\natexlab{b}})\citenamefont{Freedman, Larsen, and
  Wang}}]{Freedman02b}
\bibinfo{author}{\bibfnamefont{M.~H.} \bibnamefont{Freedman}},
  \bibinfo{author}{\bibfnamefont{M.~J.} \bibnamefont{Larsen}},
  \bibnamefont{and} \bibinfo{author}{\bibfnamefont{Z.}~\bibnamefont{Wang}},
  \bibinfo{journal}{Commun. Math. Phys.} \textbf{\bibinfo{volume}{228}},
  \bibinfo{pages}{177} (\bibinfo{year}{2002}{\natexlab{b}}),
  \eprint{math/0103200}.

\bibitem[{\citenamefont{Freedman et~al.}(2003)\citenamefont{Freedman, Kitaev,
  Larsen, and Wang}}]{Freedman03b}
\bibinfo{author}{\bibfnamefont{M.~H.} \bibnamefont{Freedman}},
  \bibinfo{author}{\bibfnamefont{A.}~\bibnamefont{Kitaev}},
  \bibinfo{author}{\bibfnamefont{M.~J.} \bibnamefont{Larsen}},
  \bibnamefont{and} \bibinfo{author}{\bibfnamefont{Z.}~\bibnamefont{Wang}},
  \bibinfo{journal}{Bull. Amer. Math. Soc.}
  \textbf{\bibinfo{volume}{40}}, \bibinfo{pages}{31} (\bibinfo{year}{2003}), \eprint{quant-ph/0101025}.

\bibitem[{\citenamefont{Nayak et~al.}(2008)\citenamefont{Nayak, Simon, Stern,
  Freedman, and Das~Sarma}}]{Nayak08}
\bibinfo{author}{\bibfnamefont{C.}~\bibnamefont{Nayak}},
  \bibinfo{author}{\bibfnamefont{S.~H.} \bibnamefont{Simon}},
  \bibinfo{author}{\bibfnamefont{A.}~\bibnamefont{Stern}},
  \bibinfo{author}{\bibfnamefont{M.}~\bibnamefont{Freedman}}, \bibnamefont{and}
  \bibinfo{author}{\bibfnamefont{S.}~\bibnamefont{Das~Sarma}},
  \bibinfo{journal}{Rev. Mod. Phys.} \textbf{\bibinfo{volume}{80}},
  \bibinfo{pages}{1083} (\bibinfo{year}{2008}), \eprint{arXiv:0707.1889}.

\bibitem[{\citenamefont{Mochon}(2003)}]{Mochon03}
\bibinfo{author}{\bibfnamefont{C.}~\bibnamefont{Mochon}},
  \bibinfo{journal}{Phys. Rev. A} \textbf{\bibinfo{volume}{67}},
  \bibinfo{eid}{022315} (\bibinfo{year}{2003}),
  \eprint{quant-ph/0206128}.

\bibitem[{\citenamefont{Mochon}(2004)}]{Mochon04}
\bibinfo{author}{\bibfnamefont{C.}~\bibnamefont{Mochon}},
  \bibinfo{journal}{Phys. Rev. A} \textbf{\bibinfo{volume}{69}},
  \bibinfo{pages}{032306} (\bibinfo{year}{2004}), \eprint{quant-ph/0306063}.

\bibitem[{\citenamefont{Bonderson
  et~al.}(2008{\natexlab{a}})\citenamefont{Bonderson, Freedman, and
  Nayak}}]{Bonderson08a}
\bibinfo{author}{\bibfnamefont{P.}~\bibnamefont{Bonderson}},
  \bibinfo{author}{\bibfnamefont{M.}~\bibnamefont{Freedman}}, \bibnamefont{and}
  \bibinfo{author}{\bibfnamefont{C.}~\bibnamefont{Nayak}},
  \bibinfo{journal}{Phys. Rev. Lett.} \textbf{\bibinfo{volume}{101}},
  \bibinfo{pages}{010501} (\bibinfo{year}{2008}{\natexlab{a}}),
  \eprint{arXiv:0802.0279}.

\bibitem[{\citenamefont{Bonderson et~al.}(2009)\citenamefont{Bonderson,
  Freedman, and Nayak}}]{Bonderson08b}
\bibinfo{author}{\bibfnamefont{P.}~\bibnamefont{Bonderson}},
  \bibinfo{author}{\bibfnamefont{M.}~\bibnamefont{Freedman}}, \bibnamefont{and}
  \bibinfo{author}{\bibfnamefont{C.}~\bibnamefont{Nayak}},
  \bibinfo{journal}{Annals Phys.} \textbf{\bibinfo{volume}{324}},
  \bibinfo{pages}{787} (\bibinfo{year}{2009}), \eprint{arXiv:0808.1933}.

\bibitem[{\citenamefont{Bonderson}(2013)}]{Bonderson12a}
\bibinfo{author}{\bibfnamefont{P.}~\bibnamefont{Bonderson}},
  \bibinfo{journal}{Phys. Rev. B} \textbf{\bibinfo{volume}{87}},
  \bibinfo{pages}{035113} (\bibinfo{year}{2013}), \eprint{arXiv:1210.7929}.

\bibitem[{\citenamefont{Bonderson
  et~al.}({\natexlab{a}})\citenamefont{Bonderson, Freedman, Lutchyn, Nayak, and
  Wang}}]{Bonderson-unpublished}
\bibinfo{author}{\bibfnamefont{P.}~\bibnamefont{Bonderson}},
  \bibinfo{author}{\bibfnamefont{M.}~\bibnamefont{Freedman}},
  \bibinfo{author}{\bibfnamefont{R.}~\bibnamefont{Lutchyn}},
  \bibinfo{author}{\bibfnamefont{C.}~\bibnamefont{Nayak}}, \bibnamefont{and}
  \bibinfo{author}{\bibfnamefont{Z.}~\bibnamefont{Wang}},
  \bibinfo{note}{unpublished}.

\bibitem[{\citenamefont{Read and Rezayi}(1999)}]{Read99}
\bibinfo{author}{\bibfnamefont{N.}~\bibnamefont{Read}} \bibnamefont{and}
  \bibinfo{author}{\bibfnamefont{E.}~\bibnamefont{Rezayi}},
  \bibinfo{journal}{Phys. Rev. B} \textbf{\bibinfo{volume}{59}},
  \bibinfo{pages}{8084} (\bibinfo{year}{1999}), \eprint{cond-mat/9809384}.

\bibitem[{\citenamefont{Hermanns}(2010)}]{Hermanns09}
\bibinfo{author}{\bibfnamefont{M.}~\bibnamefont{Hermanns}},
  \bibinfo{journal}{Phys. Rev. Lett.} \textbf{\bibinfo{volume}{104}},
  \bibinfo{pages}{056803} (\bibinfo{year}{2010}), \eprint{arXiv:0906.2073}.

\bibitem[{\citenamefont{Pan et~al.}(1999)\citenamefont{Pan, Xia, Shvarts,
  Adams, Stormer, Tsui, Pfeiffer, Baldwin, and West}}]{Pan99}
\bibinfo{author}{\bibfnamefont{W.}~\bibnamefont{Pan}},
  \bibinfo{author}{\bibfnamefont{J.-S.} \bibnamefont{Xia}},
  \bibinfo{author}{\bibfnamefont{V.}~\bibnamefont{Shvarts}},
  \bibinfo{author}{\bibfnamefont{D.~E.} \bibnamefont{Adams}},
  \bibinfo{author}{\bibfnamefont{H.~L.} \bibnamefont{Stormer}},
  \bibinfo{author}{\bibfnamefont{D.~C.} \bibnamefont{Tsui}},
  \bibinfo{author}{\bibfnamefont{L.~N.} \bibnamefont{Pfeiffer}},
  \bibinfo{author}{\bibfnamefont{K.~W.} \bibnamefont{Baldwin}},
  \bibnamefont{and} \bibinfo{author}{\bibfnamefont{K.~W.} \bibnamefont{West}},
  \bibinfo{journal}{Phys. Rev. Lett.} \textbf{\bibinfo{volume}{83}},
  \bibinfo{pages}{3530} (\bibinfo{year}{1999}), \eprint{cond-mat/9907356}.

\bibitem[{\citenamefont{Xia et~al.}(2004)\citenamefont{Xia, Pan, Vicente,
  Adams, Sullivan, Stormer, Tsui, Pfeiffer, Baldwin, and West}}]{Xia04}
\bibinfo{author}{\bibfnamefont{J.~S.} \bibnamefont{Xia}},
  \bibinfo{author}{\bibfnamefont{W.}~\bibnamefont{Pan}},
  \bibinfo{author}{\bibfnamefont{C.~L.} \bibnamefont{Vicente}},
  \bibinfo{author}{\bibfnamefont{E.~D.} \bibnamefont{Adams}},
  \bibinfo{author}{\bibfnamefont{N.~S.} \bibnamefont{Sullivan}},
  \bibinfo{author}{\bibfnamefont{H.~L.} \bibnamefont{Stormer}},
  \bibinfo{author}{\bibfnamefont{D.~C.} \bibnamefont{Tsui}},
  \bibinfo{author}{\bibfnamefont{L.~N.} \bibnamefont{Pfeiffer}},
  \bibinfo{author}{\bibfnamefont{K.~W.} \bibnamefont{Baldwin}},
  \bibnamefont{and} \bibinfo{author}{\bibfnamefont{K.~W.} \bibnamefont{West}},
  \bibinfo{journal}{Phys. Rev. Lett.} \textbf{\bibinfo{volume}{93}},
  \bibinfo{pages}{176809} (\bibinfo{year}{2004}), \eprint{cond-mat/0406724}.

\bibitem[{\citenamefont{Pan et~al.}(2008)\citenamefont{Pan, Xia, Stormer, Tsui,
  Vicente, Adams, Sullivan, Pfeiffer, Baldwin, and West}}]{Pan08}
\bibinfo{author}{\bibfnamefont{W.}~\bibnamefont{Pan}},
  \bibinfo{author}{\bibfnamefont{J.~S.} \bibnamefont{Xia}},
  \bibinfo{author}{\bibfnamefont{H.~L.} \bibnamefont{Stormer}},
  \bibinfo{author}{\bibfnamefont{D.~C.} \bibnamefont{Tsui}},
  \bibinfo{author}{\bibfnamefont{C.}~\bibnamefont{Vicente}},
  \bibinfo{author}{\bibfnamefont{E.~D.} \bibnamefont{Adams}},
  \bibinfo{author}{\bibfnamefont{N.~S.} \bibnamefont{Sullivan}},
  \bibinfo{author}{\bibfnamefont{L.~N.} \bibnamefont{Pfeiffer}},
  \bibinfo{author}{\bibfnamefont{K.~W.} \bibnamefont{Baldwin}},
  \bibnamefont{and} \bibinfo{author}{\bibfnamefont{K.~W.} \bibnamefont{West}},
  \bibinfo{journal}{Phys. Rev. B} \textbf{\bibinfo{volume}{77}},
  \bibinfo{pages}{075307} (\bibinfo{year}{2008}), \eprint{ar{X}iv:0801.1318}.

\bibitem[{\citenamefont{{Pan} et~al.}(2012)\citenamefont{{Pan}, {Baldwin},
  {West}, {Pfeiffer}, and {Tsui}}}]{Pan12}
\bibinfo{author}{\bibfnamefont{W.}~\bibnamefont{{Pan}}},
  \bibinfo{author}{\bibfnamefont{K.~W.} \bibnamefont{{Baldwin}}},
  \bibinfo{author}{\bibfnamefont{K.~W.} \bibnamefont{{West}}},
  \bibinfo{author}{\bibfnamefont{L.~N.} \bibnamefont{{Pfeiffer}}},
  \bibnamefont{and} \bibinfo{author}{\bibfnamefont{D.~C.}
  \bibnamefont{{Tsui}}}, \bibinfo{journal}{Phys. Rev. Lett.}
  \textbf{\bibinfo{volume}{108}}, \bibinfo{pages}{216804}
  (\bibinfo{year}{2012}), \eprint{arXiv:1204.0557}.

\bibitem[{\citenamefont{Cui and Wang}()}]{Cui14}
\bibinfo{author}{\bibfnamefont{S.~X.} \bibnamefont{Cui}} \bibnamefont{and}
  \bibinfo{author}{\bibfnamefont{Z.}~\bibnamefont{Wang}},
  \bibinfo{journal}{J. Math. Phys.} \textbf{\bibinfo{volume}{56}},
  \bibinfo{pages}{032202} (\bibinfo{year}{2015}),
  \eprint{arXiv:1405.7778}.

\bibitem[{\citenamefont{Kitaev}(2006)}]{Kitaev06a}
\bibinfo{author}{\bibfnamefont{A.}~\bibnamefont{Kitaev}},
  \bibinfo{journal}{Annals Phys.} \textbf{\bibinfo{volume}{321}},
  \bibinfo{pages}{2} (\bibinfo{year}{2006}), \eprint{cond-mat/0506438}.

\bibitem[{\citenamefont{Bonderson}(2007)}]{Bonderson07b}
\bibinfo{author}{\bibfnamefont{P.~H.} \bibnamefont{Bonderson}}, Ph.D. thesis,
  \bibinfo{school}{California Institute of Technology} (\bibinfo{year}{2007}).

\bibitem[{\citenamefont{Mac~Lane}(1998)}]{MacLane98}
\bibinfo{author}{\bibfnamefont{S.}~\bibnamefont{Mac~Lane}},
  \emph{\bibinfo{title}{{C}ategories for the {W}orking {M}athematician}},
  {G}raduate {T}exts in {M}athematics (\bibinfo{publisher}{Springer-Verlag},
  \bibinfo{address}{New York}, \bibinfo{year}{1998}), \bibinfo{edition}{2nd}
  ed.

\bibitem[{\citenamefont{Bonderson
  et~al.}(2008{\natexlab{b}})\citenamefont{Bonderson, Shtengel, and
  Slingerland}}]{Bonderson07c}
\bibinfo{author}{\bibfnamefont{P.}~\bibnamefont{Bonderson}},
  \bibinfo{author}{\bibfnamefont{K.}~\bibnamefont{Shtengel}}, \bibnamefont{and}
  \bibinfo{author}{\bibfnamefont{J.~K.} \bibnamefont{Slingerland}},
  \bibinfo{journal}{Annals of Physics} \textbf{\bibinfo{volume}{323}},
  \bibinfo{pages}{2709} (\bibinfo{year}{2008}{\natexlab{b}}),
  \eprint{arXiv:0707.4206}.

\bibitem[{\citenamefont{Bonderson et~al.}(2007)\citenamefont{Bonderson,
  Shtengel, and Slingerland}}]{Bonderson07a}
\bibinfo{author}{\bibfnamefont{P.}~\bibnamefont{Bonderson}},
  \bibinfo{author}{\bibfnamefont{K.}~\bibnamefont{Shtengel}}, \bibnamefont{and}
  \bibinfo{author}{\bibfnamefont{J.~K.} \bibnamefont{Slingerland}},
  \bibinfo{journal}{Phys. Rev. Lett.} \textbf{\bibinfo{volume}{98}},
  \bibinfo{pages}{070401} (\bibinfo{year}{2007}), \eprint{quant-ph/0608119}.

\bibitem[{\citenamefont{Bonderson
  et~al.}({\natexlab{b}})\citenamefont{Bonderson, Fidkowski, Freedman, and
  Walker}}]{Bonderson13b}
\bibinfo{author}{\bibfnamefont{P.}~\bibnamefont{Bonderson}},
  \bibinfo{author}{\bibfnamefont{L.}~\bibnamefont{Fidkowski}},
  \bibinfo{author}{\bibfnamefont{M.}~\bibnamefont{Freedman}}, \bibnamefont{and}
  \bibinfo{author}{\bibfnamefont{K.}~\bibnamefont{Walker}},
  \bibinfo{note}{arXiv:1306.2379}.

\bibitem[{\citenamefont{{Freedman} and {Levaillant}}()}]{Freedman15}
\bibinfo{author}{\bibfnamefont{M.~H.} \bibnamefont{{Freedman}}}
  \bibnamefont{and} \bibinfo{author}{\bibfnamefont{C.~I.}
  \bibnamefont{{Levaillant}}}, \eprint{arXiv:1501.01339}.

\bibitem[{\citenamefont{Levaillant}()}]{Levaillant15b}
\bibinfo{author}{\bibfnamefont{C.}~\bibnamefont{Levaillant}},
  \eprint{arXiv:1501.02841}.

\bibitem[{\citenamefont{Olmsted}(1945)}]{Olmsted1945}
\bibinfo{author}{\bibfnamefont{J.~M.~H.} \bibnamefont{Olmsted}},
  \bibinfo{journal}{The American Mathematical Monthly}
  \textbf{\bibinfo{volume}{52}}, \bibinfo{pages}{507} (\bibinfo{year}{1945}).

\bibitem[{\citenamefont{{Bocharov} et~al.}()\citenamefont{{Bocharov},
  {Roetteler}, and {Svore}}}]{Bocharov2014}
\bibinfo{author}{\bibfnamefont{A.}~\bibnamefont{{Bocharov}}},
  \bibinfo{author}{\bibfnamefont{M.}~\bibnamefont{{Roetteler}}},
  \bibnamefont{and} \bibinfo{author}{\bibfnamefont{K.~M.}
  \bibnamefont{{Svore}}},
  \bibinfo{journal}{Phys. Rev. A} \textbf{\bibinfo{volume}{91}},
  \bibinfo{pages}{052317} (\bibinfo{year}{2015}),
   \eprint{arXiv:1409.3552}.

\bibitem[{\citenamefont{Bocharov et~al.}()\citenamefont{Bocharov, Gurevich, and
  Svore}}]{Bocharov-unpublished}
\bibinfo{author}{\bibfnamefont{A.}~\bibnamefont{Bocharov}},
  \bibinfo{author}{\bibfnamefont{Y.}~\bibnamefont{Gurevich}}, \bibnamefont{and}
  \bibinfo{author}{\bibfnamefont{K.}~\bibnamefont{Svore}},
  \bibinfo{note}{unpublished}.

\bibitem[{\citenamefont{{Barkeshli} et~al.}()\citenamefont{{Barkeshli},
  {Bonderson}, {Cheng}, and {Wang}}}]{Barkeshli14}
\bibinfo{author}{\bibfnamefont{M.}~\bibnamefont{{Barkeshli}}},
  \bibinfo{author}{\bibfnamefont{P.}~\bibnamefont{{Bonderson}}},
  \bibinfo{author}{\bibfnamefont{M.}~\bibnamefont{{Cheng}}}, \bibnamefont{and}
  \bibinfo{author}{\bibfnamefont{Z.}~\bibnamefont{{Wang}}},
  \eprint{arXiv:1410.4540}.

\bibitem[{\citenamefont{{Barkeshli} et~al.}(2013)\citenamefont{{Barkeshli},
  {Jian}, and {Qi}}}]{Barkeshli2013}
\bibinfo{author}{\bibfnamefont{M.}~\bibnamefont{{Barkeshli}}},
  \bibinfo{author}{\bibfnamefont{C.-M.} \bibnamefont{{Jian}}},
  \bibnamefont{and} \bibinfo{author}{\bibfnamefont{X.-L.} \bibnamefont{{Qi}}},
  \bibinfo{journal}{\prb} \textbf{\bibinfo{volume}{87}},
  \bibinfo{pages}{045130} (\bibinfo{year}{2013}), \eprint{arXiv:1208.4834}.

\bibitem[{\citenamefont{Chari and Pressley}(1995)}]{Chari-book}
\bibinfo{author}{\bibfnamefont{V.}~\bibnamefont{Chari}} \bibnamefont{and}
  \bibinfo{author}{\bibfnamefont{A.~N.} \bibnamefont{Pressley}},
  \emph{\bibinfo{title}{A Guide to Quantum Groups}}
  (\bibinfo{publisher}{Cambridge University Press}, \bibinfo{year}{1995}).

\bibitem[{\citenamefont{Witten}(1989)}]{Witten89}
\bibinfo{author}{\bibfnamefont{E.}~\bibnamefont{Witten}},
  \bibinfo{journal}{Commun. Math. Phys.} \textbf{\bibinfo{volume}{121}},
  \bibinfo{pages}{351} (\bibinfo{year}{1989}).

\bibitem[{\citenamefont{Wess and Zumino}(1971)}]{Wess71}
\bibinfo{author}{\bibfnamefont{J.}~\bibnamefont{Wess}} \bibnamefont{and}
  \bibinfo{author}{\bibfnamefont{B.}~\bibnamefont{Zumino}},
  \bibinfo{journal}{Phys. Lett. B} \textbf{\bibinfo{volume}{37}},
  \bibinfo{pages}{95} (\bibinfo{year}{1971}).

\bibitem[{\citenamefont{Witten}(1983)}]{Witten83}
\bibinfo{author}{\bibfnamefont{E.}~\bibnamefont{Witten}},
  \bibinfo{journal}{Nucl. Phys. B} \textbf{\bibinfo{volume}{223}},
  \bibinfo{pages}{422} (\bibinfo{year}{1983}).

\bibitem[{\citenamefont{Jones}(1985)}]{Jones85}
\bibinfo{author}{\bibfnamefont{V.~F.~R.} \bibnamefont{Jones}},
  \bibinfo{journal}{Bull. Am. Math. Soc.} \textbf{\bibinfo{volume}{12}},
  \bibinfo{pages}{103} (\bibinfo{year}{1985}).

\bibitem[{\citenamefont{Kauffman and Lins}(1994)}]{Kauffman-book}
\bibinfo{author}{\bibfnamefont{L.~H.} \bibnamefont{Kauffman}} \bibnamefont{and}
  \bibinfo{author}{\bibfnamefont{S.}~\bibnamefont{Lins}},
  \emph{\bibinfo{title}{Temperley-Lieb Recoupling Theory and Invariants of
  3-manifolds}} (\bibinfo{publisher}{Princeton University Press},
  \bibinfo{year}{1994}).

\bibitem[{\citenamefont{Wang}(2010)}]{Wang-book}
\bibinfo{author}{\bibfnamefont{Z.}~\bibnamefont{Wang}},
  \emph{\bibinfo{title}{Topological Quantum Computation}}
  (\bibinfo{publisher}{American Mathematical Society}, \bibinfo{year}{2010}).

\end{thebibliography}

\end{document}